\documentclass[prl,twocolumn, showpacs, floatfix,tightenlines amsmath, amssymb, preprintnumbers, superscriptaddress, longbibliography] {revtex4-2}
\usepackage[pdftex]{graphicx} 

\usepackage{xcolor}
\usepackage{url}
\usepackage{amsmath}
\usepackage{float}
\usepackage[colorlinks=true, citecolor=blue, linkcolor=blue, urlcolor=blue]{hyperref}
\usepackage{multirow}
\usepackage{physics}
\usepackage{tabularx}
\usepackage{xr}

\newcommand{\NPS}{NiPS$_3$} 
\newcommand{\Nid}{Ni$^{2+}$}
\newcommand{\eloss}{$E_{\mathrm{loss}}$}
\newcommand{\elossR}[2]{\eloss{} = #1 $\rightarrow$ #2 eV}
\newcommand{\Si}{SiO$_2$}
\newcommand{\dCT}[2]{$3d^{#1}\underline{L}^{#2}$} 
\newcommand{\de}{$\ket{3d^8}$}
\newcommand{\dn}{$\ket{3d^9\underline{L}^1}$}
\newcommand{\Dq}{10\textit{Dq}}

\newcommand{\pds}{$pd\sigma$}
\newcommand{\pdp}{$pd\pi$}
\newcommand{\ppsi}{$pp\sigma$}
\newcommand{\pppi}{$pp\pi$}
\newcommand{\Tpp}{$T_{pp}$}
\newcommand{\Tpd}{$T_{pd}$}
\newcommand{\CTChr}{$\Delta/U$}
\newcommand{\HybChr}{$\Delta/T_{pd}$}

\newcommand{\FrmtSym}[3]{${}^{#1}#2_{#3}$}
\newcommand{\DEx}[2]{$\Delta E_{#1}^{#2}$}
\newcommand{\iNN}[1]{${}^{#1}$NN}
\newcommand{\pir}[2]{$p^{#1}_{#2}$}
\newcommand{\Lxy}{$\frac{1}{2}$(\pir{1}{x} $-$ \pir{2}{y} $+$ \pir{3}{y} $-$ \pir{4}{x})}
\newcommand{\Lz}{$\frac{1}{\sqrt{3}}$(\pir{5}{z} $-$ \pir{6}{z}) $+$ $\frac{1}{2\sqrt{3}}($\pir{1}{x} $+$ \pir{2}{y} $-$ \pir{3}{y} $-$ \pir{4}{x})}
\newcommand{\Jide}[1]{$J_{#1}^{\alpha}$}
\newcommand{\Jidn}[1]{$J_{#1}^{\beta}$}
\newcommand{\FigSumm}{Fig.~\ref{fig:Results_FittingSummary}}
\newcommand{\FigTS}{Fig.~\ref{fig:BulkTS}}

\newcommand{\FigGS}{Fig.~\ref{fig:Modeling_GS-SE}}
\newcommand{\ab}{\textit{ab initio}}
\newcommand{\NiS}{NiS$_6$}
\newcommand{\trig}{$D_{3d}$}
\newcommand{\Oh}{$O_h$}
\newcommand{\Fdd}{$F_{dd}^{0,2,4}$}
\newcommand{\um}{$\mu m$}

\begin{document}
\title{Dimensionality dependent electronic structure of the exfoliated van der Waals antiferromagnet \NPS{}}
\author{M. F. DiScala}
\affiliation{Department of Physics, Brown University, Providence, RI 02912}

\author{D. Staros}
\affiliation{Department of Chemistry, Brown University, Providence, RI 02912}
\author{A. de la Torre}
\affiliation{Department of Physics, Brown University, Providence, RI 02912}

\author{A. Lopez}
\affiliation{Department of Physics, Brown University, Providence, RI 02912}
\author{D. Wong}
\affiliation{Department of Dynamics and Transport in Quantum Materials, Helmholtz-Zentrum Berlin
für Materialen und Energie, Albert-Einstein-Strasse 15, 12489 Berlin,
Germany}
\author{C. Schulz}
\affiliation{Department of Dynamics and Transport in Quantum Materials, Helmholtz-Zentrum Berlin
für Materialen und Energie, Albert-Einstein-Strasse 15, 12489 Berlin,
Germany}
\author{M. Bartkowiak}
\affiliation{Department of Dynamics and Transport in Quantum Materials, Helmholtz-Zentrum Berlin
für Materialen und Energie, Albert-Einstein-Strasse 15, 12489 Berlin,
Germany}
\author{B. Rubenstein}
\affiliation{Department of Chemistry, Brown University, Providence, RI 02912}
\author{K. W. Plumb}
\affiliation{Department of Physics, Brown University, Providence, RI 02912}
\email{Authors to whom correspondence should be addressed: Brenda Rubenstein, brenda\_rubenstein@brown.edu and Kemp Plumb, kemp\_plumb@brown.edu}
\date{\today}

\begin{abstract}
Resonant Inelastic X-ray Scattering (RIXS) was used to measure the local electronic structure in few-layer exfoliated flakes of the van der Waals antiferromagnet \NPS{}. The resulting spectra show a systematic softening and broadening of \NiS{} multiplet excitations with decreasing layer count from the bulk to three atomic layers (3L). These trends are driven by a decrease in the transition metal-ligand and ligand-ligand hopping integrals, and in the charge-transfer energy: $\Delta$ = 0.60 eV in the bulk and 0.22 eV in 3L \NPS{}. Relevant intralayer magnetic exchange integrals computed from the electronic parameters exhibit a systematic decrease in the average interaction strength with thickness and place 2D \NPS{} close to the phase boundary between stripy and spiral antiferromagnetic order, which may explain the apparent vanishing of long-range order in the 2D limit. This study explicitly demonstrates the influence of \emph{inter}layer electronic interactions on \emph{intra}layer ones in insulating magnets. As a consequence, the magnetic Hamiltonian in few-layer insulating magnets can be significantly different from that in the bulk. 
\end{abstract}
\maketitle

In two-dimensional magnets, enhanced fluctuations and lattice connectivity strike a balance from which collective states unobtainable in three dimensions may emerge. The ability to prepare isolated monolayers of van der Waals (vdW) magnets has enabled access to new magnetic phases and tests of fundamental theorems of magnetism \cite{parkOpportunitiesChallenges2D2016, kuoExfoliationRamanSpectroscopic2016, kimSuppressionMagneticOrdering2019, sivadasStackingDependentMagnetismBilayer2018, gongDiscoveryIntrinsicFerromagnetism2017, tangEvidenceFrustratedMagnetic2023}, and opens up possibilities for controlling or engineering unconventional states through stacking \cite{hellmanInterfaceinducedPhenomenaMagnetism2017}. Much work has concentrated on vdW materials with a net ferromagnetism in the two-dimensional (2D) limit \cite{gongDiscoveryIntrinsicFerromagnetism2017, huangLayerdependentFerromagnetismVan2017, burchMagnetismTwodimensionalVan2018, gibertiniMagnetic2DMaterials2019}. However, antiferromagnets may offer more possibilities to explore complex magnetic order, topological spin textures, or quantum spin-liquids that arise from frustrated interactions and are stabilized in 2D \cite{kitaevAnyonsExactlySolved2006, plumbRuClSpinorbitAssisted2014, nasuFermionicResponseFractionalization2016, leeMagnonicQuantumSpin2018, zhouPossibleStructuralTransformation2019}.  

\begin{figure}[t!]
  \setlength{\belowcaptionskip}{-18pt}
  \setlength{\abovecaptionskip}{-5pt}
  \begin{center}
    \includegraphics[]{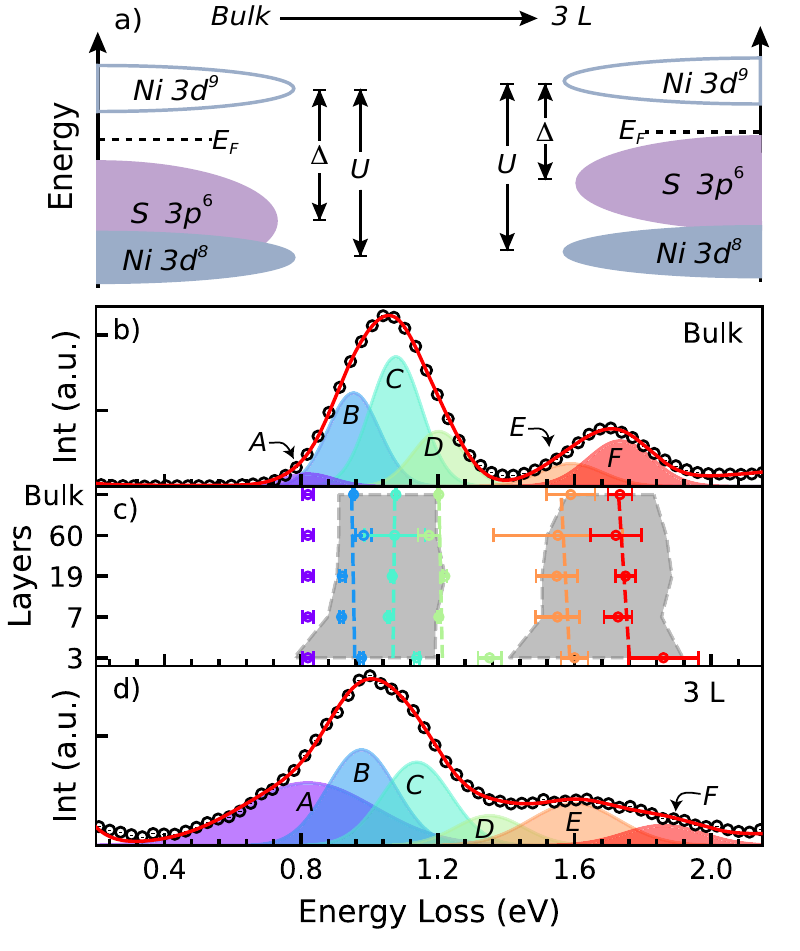}
  \caption{(a) Schematic electronic density of states from bulk to 3L \NPS{}. Ni L$_3$  RIXS spectra for bulk (b) and 3L (d) \NPS{}.  Black points are experimental data with errorbars smaller than the symbol size, red lines show Gaussian fit to the data. (c) Gaussian peak position  versus layer count. Dashed linear fits highlight the electronic structure change with thickness; gray regions follow the numerical center of mass and full width half max of the overall experimental peaks.} 
  \label{fig:Results_FittingSummary}
  \end{center}
\end{figure}

\NPS{} stands as one of the few exfoliatable materials that exhibits both antiferromagnetic order and strong correlations \cite{wildesMagneticStructureQuasitwodimensional2015, kimChargeSpinCorrelationVan2018}. Recent Raman scattering measurements suggest the magnetic order in \NPS{} is highly sensitive to dimensionality and find that long-range order vanishes in the monolayer limit in favor of a fluctuating magnetic phase \cite{kimSuppressionMagneticOrdering2019}. Based on the magnetic Hamiltonian that was determined by inelastic neutron scattering on bulk samples \cite{lanconMagneticExchangeParameters2018}, the thickness-dependent Raman data were associated with the proliferation of vorticies through a Berezinskii–Kosterlitz–Thoules phase transition in the 2D material. However, this explanation assumes that the few-layer magnetic Hamiltonian is identical to that in the bulk. More direct experimental access to the electronic energy scales and magnetic interactions is necessary to resolve the nature of the magnetic state in exfoliated \NPS{}. 

In this letter, we show that Ni-S electronic energy scales are strongly altered by dimensionality in \NPS{} and thus, the few-layer magnetic Hamiltonian differs from that of the bulk. Resonant Inelastic X-ray Scattering (RIXS) on exfoliated flakes reveals a systematic softening and broadening of \NiS{} multiplet excitations with decreasing thickness that is reproduced by a multiplet ligand-field model. Decreased hopping integrals and charge transfer energy in 2D result in a more covalent character [Fig.~\ref{fig:Results_FittingSummary}(a)]. We compute the relevant magnetic exchange integrals and find a systematic decrease in the second- and third-nearest neighbor magnetic interaction strengths, and an increase in the first-nearest neighbor interaction strength. This change moves \NPS{} closer to the boundary between the stripy antiferromagnetic and spiral ordered phases of the honeycomb antiferromagnet. The change of electronic energy scales in thinner samples occurs due to decreased electronic vdW delocalization across layers in the 2D limit. Since this mechanism is not specific to \NPS{}, it's effect will be important to the properties of a broad class of few-layer insulating vdW magnets.

\begin{figure}[t!]
  \setlength{\belowcaptionskip}{-20pt}
  \setlength{\abovecaptionskip}{-5pt}
  \begin{center}
    \includegraphics[]{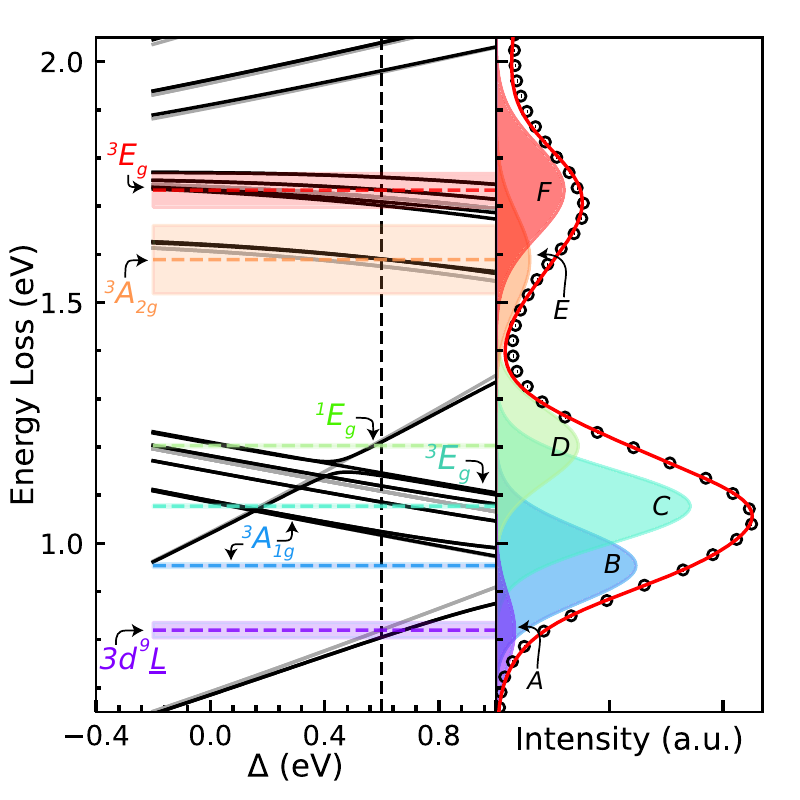}
  \caption{Calculated energy levels of the \NiS{} multiplet ligand-field model as a function of $\Delta$ (a) compared to experimental bulk spectrum (b). Fixed model parameters are listed in Table~\ref{tab:ModelParams} and Ref.\cite{Supp}. In (a), black and gray lines show energy levels calculated with and without SOC respectively. Symmetry labels adapted from a calculation without SOC. Horizontal dashed lines and shaded regions show \eloss{} value of fitted peaks in rightmost panel. Vertical dashed line indicates best fit at $\Delta$ = 0.60~eV.
  } 
  \label{fig:BulkTS}
  \end{center}
\end{figure}
Single crystals of \NPS{} were grown via standard vapor transport methods \cite{wildesMagneticStructureQuasitwodimensional2015, lanconMagneticExchangeParameters2018}. Bulk \NPS{} was exfoliated in air using conventional scotch-tape methods \cite{kuoExfoliationRamanSpectroscopic2016} and deposited either onto a blank \Si{} substrate or onto a \Si{} substrate pre-treated with a patterned Cu grid. Samples deposited onto blank \Si{} substrates were later patterned with Cu fiducial markers using electron-beam lithography \cite{Supp}. Exfoliated samples were spin coated with a PMMA protective layer and stored in an Ar atmosphere to prevent degradation. The PMMA coatings were removed immediately prior to loading the samples into the RIXS vacuum chamber via washing with acetone and isopropyl alcohol. Room temperature RIXS measurements were carried out on the PEAXIS beamline at BESSY II \cite{schulzCharacterizationSoftXray2020}. A horizontal scattering geometry of $2\theta\!=\!90^{\circ}$ was used with an $\approx$ 235~meV energy resolution (full width at half max, FWHM) using linear horizontal polarization and specular geometry. Spectra were collected in 30 minute segments to minimize sample exposure to the X-ray beam. Bulk \NPS{} spectra were collected with an identical scattering geometry, but with an $\approx$ 177~meV overall energy resolution FWHM.
 
Figs.~\ref{fig:Results_FittingSummary}(a)~\&~(b) show representative RIXS spectra for bulk and three-layer (3L) \NPS{}, respectively. Spectra were collected at the peak of the Ni L$_3$-edge XAS $E_{i}\! =\! 853$~eV, corresponding to 2$p_{3/2}$ to 3$d$ electronic transitions. We concentrate on the low energy region \elossR{0.2}{2.15} that contains excitations within the \NiS{} multiplet.  The bulk and 3-layer (3L) spectra are qualitatively similar except for a systematic overall energy broadening and softening that is readily visible in the 3L data [Fig.~\ref{fig:Results_FittingSummary} (b)]. While the qualitative similarity between bulk and 3L spectra is consistent with the fact that there are no drastic structural reconstructions upon exfoliation, the apparent broadening and softening indicates a change in the electronic structure of \NPS{} with thickness. 

In order to elucidate the origin of this change, we first concentrate our analysis on the bulk spectra and identify all relevant features. We found that six resolution-limited Gaussian modes were required to fit the bulk data, labeled \textit{A} - \textit{F} [\FigSumm{}(b)]. Each of these features can be identified as an excitation within the electronic mulitplet on the slightly trigonally distorted \NiS{} octahedra. The center of mass positions of the two bulk peaks are assigned to the $t_{2g} \rightarrow e_g$ (\textit{d-d}) excitations of \FrmtSym{3}{T}{2g} and \FrmtSym{3}{T}{1g} symmetry, respectively, in good agreement with optical measurements \cite{kimChargeSpinCorrelationVan2018, afanasievControllingAnisotropyVan2021}. The trigonal distortion introduces a $D_{3d}$ symmetry which splits \FrmtSym{3}{T}{2g} $\rightarrow$ \FrmtSym{3}{A}{1g} $+$ \FrmtSym{3}{E}{g}  (peaks \textit{B} \& \textit{C}) and \FrmtSym{3}{T}{1g} $\rightarrow$ \FrmtSym{3}{A}{2g} $+$ \FrmtSym{3}{E}{g} (peaks \textit{E} \& \textit{F}), in agreement with Raman and optical measurements \cite{afanasievControllingAnisotropyVan2021, wangElectronicRamanScattering2022}. Access to spin-flip ($\Delta S \neq 0$) excitations in the RIXS cross-section leads us to assign peak \textit{D} \FrmtSym{1}{E}{g} symmetry as the next highest excited state above \FrmtSym{3}{T}{2g} in a $3d^8$ system. Peak \textit{A} is assigned to a charge transfer excitation with \dCT{9}{1} character, where $\underline{L}^n$ denotes \textit{n} ligand holes. The 800~meV energy scale of this peak indicates a small charge transfer energy in \NPS{}. We verify these peak assignments through the application of the \NiS{} multiplet ligand-field model described below. We note that since our incident energy was tuned to the peak of the Ni L$_3$-edge XAS, our measurements were not sensitive to the sharp 1.47~eV peak reported in Ref.~\cite{kangCoherentManybodyExciton2020}.

\begin{figure}[t]
  \setlength{\belowcaptionskip}{-22pt}
  \setlength{\abovecaptionskip}{-3pt}
  \begin{center}
    \includegraphics[]{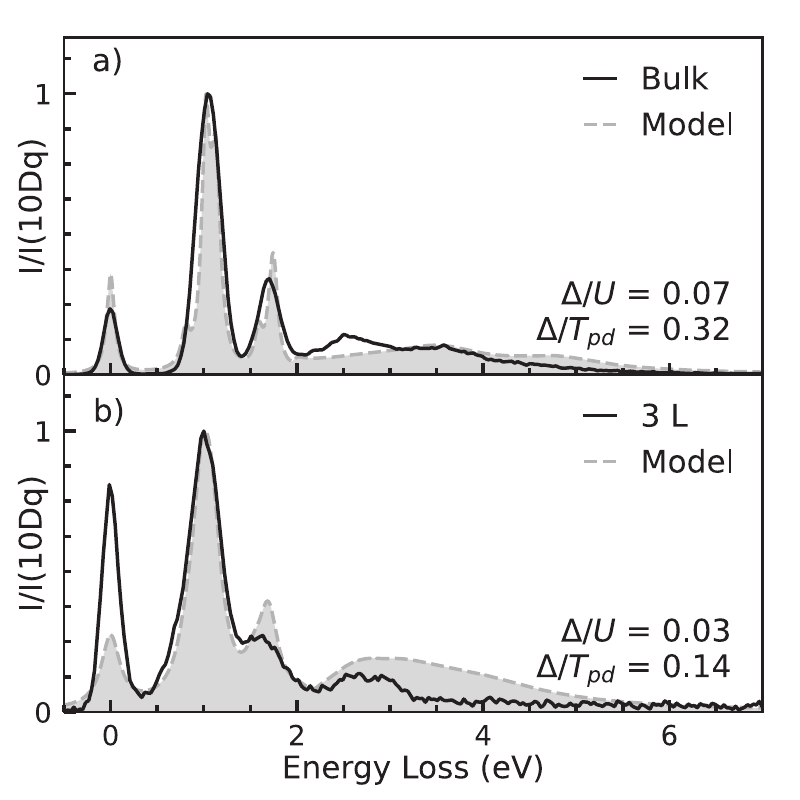}
  \caption{(a) \& (b) Normalized Ni L$_3$ RIXS spectra for bulk (a) and 3L (c) \NPS{} with AIM model from parameters in Table~\ref{tab:ModelParams} and Ref.\cite{Supp}. (c) \& (d) Simulated RIXS map as a function of incident energy from resonance (\DEx{i}{Res}) and \eloss{}. \HybChr{} and \CTChr{} parameterize the hybridization and charge transfer characters respectively.} 
  \label{fig:Results_ModelingSummary}
  \end{center}
\end{figure}
In order to facilitate an efficient exploration of the parameter space, we utilize a basis of symmetry-adapted linear combinations of ligand orbitals \cite{haverkortMultipletLigandfieldTheory2012} within a multiplet ligand-field model. Our model includes Slater-Condon parameters ($F^{0,2,4}_{dd}$, $F^{0,2}_{pd}$, $G^{1,3}_{pd}$), covalent hopping integrals between S 3\textit{p}- and Ni 3\textit{d}-orbitals, \pds{} and \pdp{}, S 3\textit{p}- orbital level splitting, \Tpp{} = \ppsi{} $-$ \pppi{}, cubic crystal field (\Dq{}) and trigonal distortion $\delta$, Ni $3d$-$3d$ and $2p$-$3d$ on-site Coulomb interactions $U_{dd}$ and $U_{pd}\!=\!1.2 U_{dd}$, and charge transfer energy $\Delta$. For initial comparisons of this model to our data, we used physically meaningful parameters for octahedrally coordinated \NiS{} \cite{bocquetElectronicStructureTransitionmetal1992,krishnakumarXrayPhotoemissionStudy2003, takuboUnusualSuperexchangePathways2007}. We carried out a search of the parameter space for $\Delta$, \Dq{}, and $F^{0,2,4}_{dd}$ by  minimizing the difference between calculated energies peaks \textit{A} - \textit{F} while keeping $F(G)_{pd}$ fixed to 80\% of their atomic Hartree-Fock values \cite{Supp,ghiringhelliNiOTestCase2005}. \FigTS{} shows the calculated energy levels for a \NiS{} cluster as a function of the charge transfer energy $\Delta$ for fixed parameters that give the best agreement between measured and calculated peak energies \cite{Supp}. Previous optical and X-ray absorption (XAS) studies classified \NPS{} as a negative charge transfer insulator \cite{kimChargeSpinCorrelationVan2018}, while more recent RIXS and XAS measurements indicate a positive charge transfer gap \cite{kangCoherentManybodyExciton2020, yanCorrelationsElectronicStructure2021}. We find that a small positive $\Delta\!=\!0.60$~eV was necessary to give an accurate match to the data.

We now bring our attention the 3L sample. Empirically fitting the 3L spectra to a minimum of six Gaussian peaks resulted in two scenarios of equally good fit quality. In scenario one, the widths of all peaks were held fixed at the experimental resolution; this fit converged with a systematic softening of all peaks between the bulk and 3L data sets. In scenario two, all peak widths were allowed to relax; this fit converged with minimal softening of all peaks, but systematic broadening and increased spectral weight attributed to peak \textit{A}. We found that peak energies extracted from scenario one could only be reproduced within physically meaningful parameters using a \emph{negative charge transfer energy} while scenario two is reproduced with a small positive charge transfer energy \cite{Supp}. A negative charge transfer energy for the 3L sample implies a zero-crossing of the charge transfer energy as a function of thickness between bulk and 2D exfoliated samples. We rule out such a transition based on the smooth evolution of thickness dependent RIXS data and Raman spectra \cite{wangElectronicRamanScattering2022}. 

Fits for scenario two are shown in \FigSumm{}(d), while \FigSumm{}(c) summarizes the centroids of the fitted peaks for the various sample thicknesses measured. Minimal differences were found between the bulk and 60L, placing a lower limit on bulk behavior for exfoliated \NPS{} at $\approx$ 38~nm. From bulk to 3L, the \FrmtSym{3}{T}{2g} modes (\FrmtSym{3}{A}{1g} + \FrmtSym{3}{E}{g}) shift slightly upward in \eloss{}, while \FrmtSym{1}{E}{g} has the largest upward shift of \DEx{loss}{} = +149(24)~meV; the \FrmtSym{3}{T}{1g} modes (\FrmtSym{3}{A}{2g} + \FrmtSym{3}{E}{g}) also shift in \eloss{}. However, both the \FrmtSym{3}{A}{2g} and \FrmtSym{3}{E}{g} modes become mixed with higher energy \FrmtSym{1}{T}{2g}(${}^1D$) modes, split by $D_{3d}$ symmetry, in few-layer samples \cite{Supp}. We find a 161(10)\% increase in the FWHM of peak \textit{A} over the bulk data, suggesting a change in the charge transfer energy in few-layer samples. This observed systematic softening and broadening of excitations signifies an electronic structure intricately connected to sample thickness in \NPS{}.

\setlength{\abovecaptionskip}{-5pt}
\begin{table}[b]
\caption{Fixed values in eV of hopping integrals extracted from \ab{} calculations of the nonmagnetic configuration. Charge transfer energy $\Delta$, and intra-orbital Coulomb repulsion $U = F^0_{dd} + \frac{4}{49}(F^2_{dd} + F^4_{dd})$ extracted from RIXS modeling \cite{haverkortMultipletLigandfieldTheory2012, shenRoleOxygenStates2022}.}
\label{tab:ModelParams}
\begin{tabularx}{\the\columnwidth}{ >{\centering\arraybackslash}X 
>{\centering\arraybackslash}X 
>{\centering\arraybackslash}X 
>{\centering\arraybackslash}X 
>{\centering\arraybackslash}X 
>{\centering\arraybackslash}X 
>{\centering\arraybackslash}X
>{\centering\arraybackslash}X }
\hline \hline
 & \pds{} & \pdp{} & \ppsi{} & \pppi{} & \Tpp{} & $\Delta$ & \textit{U} \\ \hline 
Bulk & -1.07 & 0.67 & 0.89 & -0.09 & 0.98 & 0.60 & 8.3\\
3L & -0.93 & 0.46 & 0.62 & -0.01 & 0.63 & 0.22 & 8.3 \\ \hline \hline
\end{tabularx}
\end{table}
Having identified a clear empirical trend, we then turned to \ab{} calculations for further insight. We used Density Functional Theory (DFT) to converge the electronic ground state of \NPS{} in both the bulk and monolayer geometries, generated maximally-localized Wannier functions (MLWF) which spanned the ground state DFT subspace, and used the corresponding tight-binding energy cross-terms to solve for the hopping integrals within the two-center approximation \cite{Supp}. These calculations reduced ambiguities in our parameter assignments by directly providing physically-grounded constraints on our fits. In Figs.~\ref{fig:Results_ModelingSummary}~(a)~\&~(b), we show RIXS spectra calculated using the open-source toolkit EDRIXS \cite{wangEDRIXSOpenSource2019} compared to the experimental data. Intensities were normalized to the nominal \Dq{} line. Covalent hopping integrals, \pds{} and \pdp{}, as well as \Tpp{} were fixed to those obtained from \ab{} calculations of the nonmagnetic configuration [Table~\ref{tab:ModelParams}] while $\Delta$ was allowed to vary. To account for broadening of excitations not captured by our multiplet ligand-field model and facilitate better comparison with experimental data, the calculated spectra were broadened by increasing the final-state lifetime above $2$~eV in \eloss{}. We can reproduce the observed broadening and softening of \NiS{} multiplet excitations with thickness by a decrease in charge transfer energy and transition metal-ligand hopping integrals, as parameterized by \HybChr{}, and \CTChr{}, where \Tpd{} = $-\sqrt{3}pd\sigma$ and on-site 3\textit{d} Coulomb repulsion  $U = F^0_{dd} + \frac{4}{49}(F^2_{dd} + F^4_{dd})$ \cite{haverkortMultipletLigandfieldTheory2012, shenRoleOxygenStates2022}: \HybChr{} = 0.32 and \CTChr{} = 0.07 in bulk, and \HybChr{} = 0.14 and \CTChr{} = 0.03 in 3L. 



\begin{figure}[t]
  \setlength{\belowcaptionskip}{-18pt}
  \begin{center}
    \includegraphics[]{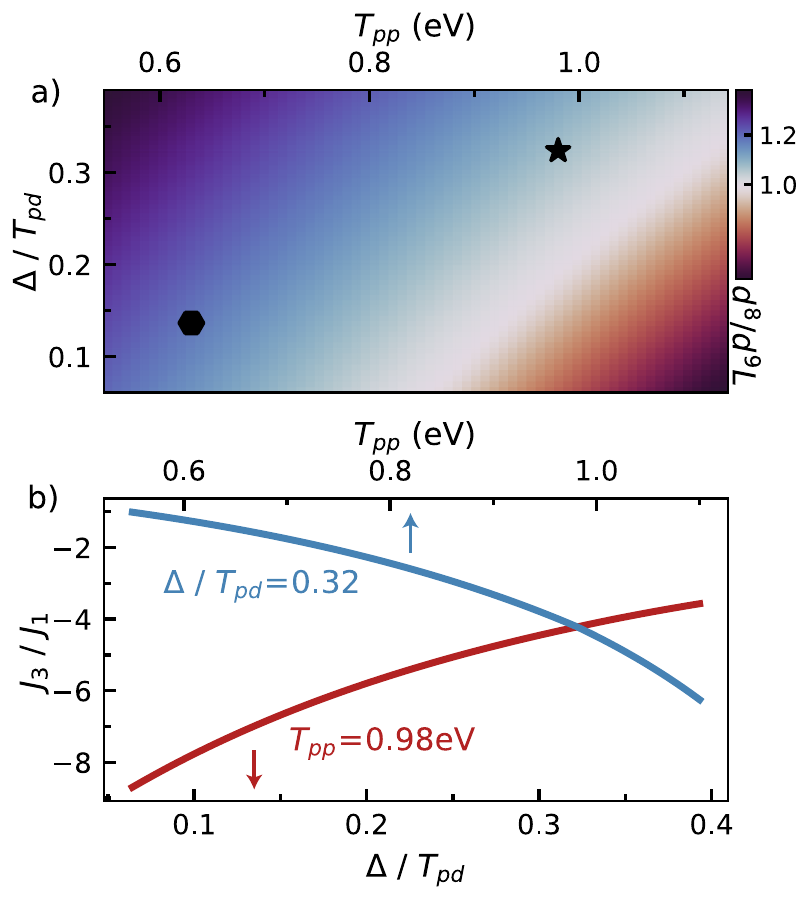}
  \caption{(a) Calculated ground state ratio of \de{}/\dn{} as a function of \Tpp{} and \HybChr{}. Star and hexagon points indicate bulk and 3L values of \Tpp{} and \HybChr{} respectively. (b) Calculated change in $J_3/J_1$ as function of \Tpp{} (top-axis) and \HybChr{} (bottom-axis) for fixed bulk values of \HybChr{} and \Tpp{}, respectively. }
  \label{fig:Modeling_GS-SE}
  \end{center}
\end{figure}

We determine that the underlying mechanism responsible for the significant change in the RIXS signal with thickness is predominantly electronic rather than structural in origin, though the lattice constant is slightly overestimated in the PBE-optimized monolayer. We find that as \NPS{} gets thinner, metal-ligand $\pi$-hopping is reduced (\pdp{} decreases) due to the removal of stabilizing, $\pi$-like, interlayer vdW interactions. The same effect also causes \pds{} and \Tpp{} to change significantly because of the mixed $\sigma$- and $\pi$-bonding character present in the $sp_3$-hybridized phosphorus atoms that bridge the \NiS{} clusters. In the context of our MLWFs, this is reflected in a change in the largest tight-binding energies used to solve for \pds{} and \pdp{} \cite{Supp}.
 
The combination of RIXS measurements and \ab{} calculations constrain the electronic ground state  that underlies the magnetic properties of \NPS{}. In \FigGS{}(a), we investigate the change in the ground state character of $\ket{\Psi_g} = \alpha\ket{3d^8} + \beta \ket{3d^9\underline{L}^1} + \gamma \ket{3d^{10}\underline{L}^2}$ extracted from our multiplet ligand-field model as a function of \Tpp{} and \HybChr{}. We find a negligible contribution from the $\ket{3d^{10}\underline{L}^2}$ state  and nearly equal populations of the \de{} and \dn{} states.  In bulk, $\abs{\alpha}^2/\,\abs{\beta}^2 = 1.08$, and in 3L, $\abs{\alpha}^2/\,\abs{\beta}^2 = 1.2$, implying a small increase in the magnitude of the paramagnetic Ni moment. As shown in \FigGS{}(a), this minimal change is accounted for by the dependence of the ground state character on both the hybridization \HybChr{} and \Tpp{}. While an increased transition metal-ligand hybridization tends to enhance the $\ket{3d^9\underline{L}^1}$ character, this is offset by the reduction in the ligand-ligand hybridization $T_{pp}$. 
 
Despite the small change in ground state character, the large change in hybridization and hopping parameters influences the magnetic exchange interactions. Using the parameters obtained from our RIXS measurements and \ab{} modeling, we compute the superexchange interactions up to the third-nearest neighbor within a sixth order cell-perturbation \cite{takuboUnusualSuperexchangePathways2007,jeffersonDerivationSinglebandModel1992, eskesSuperexchangeCuprates1993}. The first $J_1$, second $J_2$, and third $J_3$ nearest neighbor expressions are given by the second order perturbation terms for the \dn{} states, and fourth and sixth order terms for the \de{} states \cite{takuboUnusualSuperexchangePathways2007}; a detailed description of these expressions is given in Ref.~\cite{Supp}. In bulk \NPS{}, we find $J_1^{B}\!\sim\!-4.0$~meV, $J_2^{B}\!\sim\!0.25$~meV, and $J_3^{B}\!\sim\!17$~meV, in excellent agreement with recently reported values from inelastic neutron scattering \cite{wildesMagneticDynamicsMathrmNiPS2022,lanconMagneticExchangeParameters2018}. The decrease in \Tpp{} and \HybChr{} leads to an overall enhancement of $J_1^{3L}\!\sim\!-4.5$~meV,  a vanishing $J^{3L}_{2}$, and decrease in $J_3^{3L}\!\sim\!10$~meV. In \FigGS{}(b), we summarize the dependence of $J_3/J_1$ on \Tpp{} and \HybChr{}. We find that $J_1$ is dominated by the \dn{} state, while $J_{2}$ and $J_{3}$ are dominated by the \de{} state. Thus, the decrease in \Tpp{} is directly responsible for an increased \dn{} contribution to $J_1$. As a consequence of the overall reduction in the \emph{average} exchange interaction strength, the magnetic transition temperature is expected to be reduced in few-layer samples compared to bulk samples. Furthermore, the decrease in $J_3/J_1$ from -4.2 in bulk to -2.2 in 3L, positions 3L \NPS{} closer to a phase boundary between the stripy AFM phase and a spiral ordered phase \cite{fouetInvestigationQuantumModel2001}. It is likely that, in the 2D limit, \NPS{} is driven into a highly frustrated regime on this phase boundary. 

In summary, we used RIXS to access the electronic ground state properties of an exfoliated, correlated antiferromagnet in the 2D limit. We found that electronic energy scales associated with Ni-S hybridization, and consequently the magnetic exchange interactions, are altered in a non-trivial way though the modification of interlayer energy scales upon exfoliation of \NPS{} despite minimal structural changes. Our findings demonstrate that magnetic exchange parameters determined from measurements on bulk materials are not applicable in the 2D limit, as interlayer interactions, absent in 2D, affect intralayer ones. The underlying electronic mechanism we have identified points to the possibility of controlling magnetic interactions in strongly correlated van der Waals heterostuctures by tuning interfacial energy scales towards the design of the next generation of 2D strongly correlated magnetic materials. 

\section{Acknowledgments}
\begin{acknowledgments}
We thank Mark Dean for helpful discussions and comments on this manuscript. We also thank Naiyuan J. Zhang and Erin Morissette for their guidance and consultation on pattern fabrication. M.F.D. and K.W.P. were supported by the National Science Foundation under grant NO. OMA-1936221. A.D.L.T. is supported by the U.S. Department of Energy, Office of Science, Office of Basic Energy Sciences, under Award Number DE-SC0021. D.S. and B.R. were supported by the U.S. Department of Energy, Office of Science, Basic Energy Sciences, Materials Sciences and Engineering Division, as part of the Computational Materials Sciences Program and the Center for Predictive Simulation of Functional Materials, while A.L. was supported by the Brown University Diversity Fellowship. RIXS measurements were carried out at the U41-PEAXIS beamline at the BESSY II electron storage ring operated by the Helmholtz-Zentrum Berlin für Materialien und Energie. This research was conducted using computational resources and services at the Center for Computation and Visualization, Brown University. 
\end{acknowledgments}


%


\pagebreak
\clearpage
\begin{widetext}
\begin{center}
\textbf{\large Supplementary Information for "Dimensionality dependent electronic structure of the exfoliated van der Waals antiferromagnet \NPS{}"}
\end{center}
\end{widetext}

\section{Sample preparation}
\subsection{Growth and Fabrication}
Single crystal samples of \NPS{} were grown by vapor transport, following previously published methods \cite{wildesMagneticStructureQuasitwodimensional2015, lanconMagneticExchangeParameters2018}. Furnace temperature settings and duration are shown in Table~\ref{Growth}.

\renewcommand{\thetable}{SI}
\begin{table}[h!]
\caption{Furnace temperature settings and duration}
\label{Growth}
\begin{tabularx}{\the\columnwidth}{ 
>{\centering\arraybackslash}X 
>{\centering\arraybackslash}X 
>{\centering\arraybackslash}X }
\hline \hline
\; Zone 1 (${}^{\circ}$C) \; & Zone 2 (${}^{\circ}$C) & \; Duration (Days) \; \\ \hline
700 & 750 & 2 \\ 
670 & 620 & 16 \\ \hline \hline
\end{tabularx}
\end{table}

Bulk \NPS{} was exfoliated using conventional scotch-tape methods \cite{kuoExfoliationRamanSpectroscopic2016} and deposited either onto a blank \Si{} substrate or onto a \Si{} substrate pre-treated with a patterned Copper (Cu) grid. The patterned Cu grid consisted of 100 \um{} x \um{} \Si{} cells separated by 200 \um{} of 50 \textit{nm} thickness Cu [Fig.~\ref{EbeamRef}(b)~\&~(c)]. The 3L sample was deposited onto a blank \Si{} substrate and was later patterned with a Cu fiducial marker using electron-beam lithography [Fig.~\ref{EbeamRef}(a)], again with a Cu thickness of 50 \textit{nm}. In both cases, Cu was chosen as a material that could provide a fluorescence contrast to \Si{} in the soft X-ray regime. This fluorescence contrast proved invaluable in locating small samples whose signals were weak under an X-ray beam. The Cu grid proved a useful method for locating sample(s) as a unique grid scheme could be defined for each chip if the orientation of each chip remained consistent; however, the fiducial marker had the advantage of of being visible by eye, resulting in unequivocal sample location and removing the requirement of a grid scheme.

\renewcommand{\thefigure}{SI}
\begin{figure}[t]
  \setlength{\belowcaptionskip}{-18pt}
  \begin{center}
    \includegraphics[width = 1\linewidth]{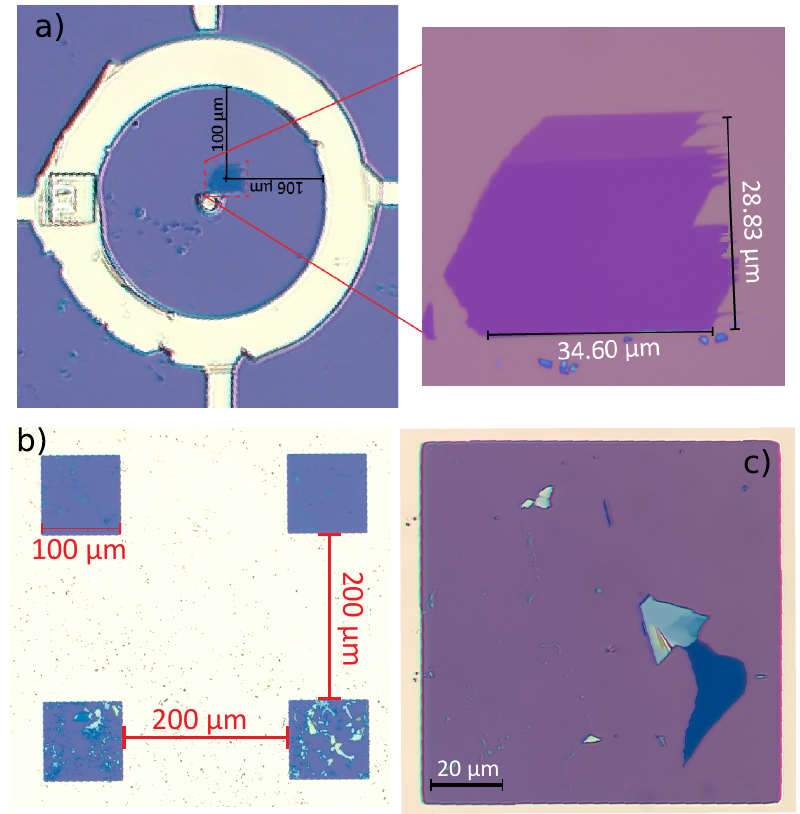}
  \caption{(a) Cu fiducial marker patterned around pre-exfoliated 3L flake of \NPS{}. Optical contrast of x100 magnification optical image was crucial in determining sample thickness. (b) Sample image of exfoliated flakes on \Si{} substrate pre-treated with patterned Cu grid. (c) x100 magnification optical image of Cu grid cell depicting measured 7L sample and dimensions.} 
  \label{EbeamRef}
  \end{center}
\end{figure}

\subsection{Sample Preservation}
\renewcommand{\thefigure}{SII}
\begin{figure}[t]
  \setlength{\belowcaptionskip}{-22pt}
   \setlength{\abovecaptionskip}{1pt}
  \begin{center}
    \includegraphics[width = 1\linewidth]{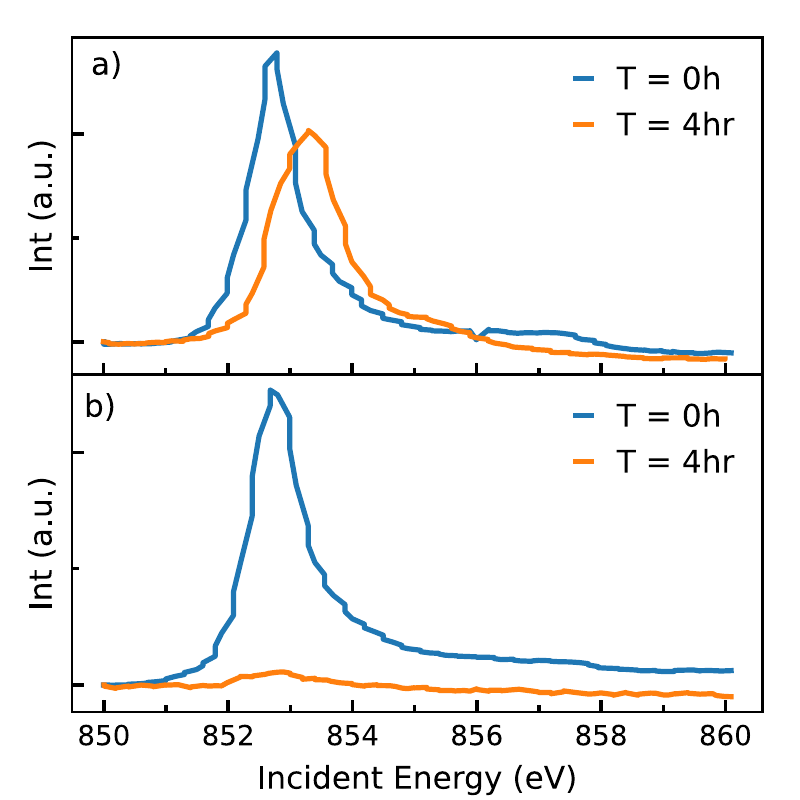}
  \caption{XAS spectra taken before and after 4 hours of beam exposure for two $\lesssim$ 5L samples of \NPS{}. (a) Sample preserved with PMMA immediately after exfoliation. (b) Sample left in air for one week prior to measurements.} 
  \label{SampleDegXAS}
  \end{center}
\end{figure}
Sample degradation in air through oxidation is a known issue for exfoliated flakes \cite{kuoExfoliationRamanSpectroscopic2016,luExfoliationLatticeVibration2020}. To ensure our samples remained intact, \Si{} chips were spin-coated with a layer of PMMA immediately after exfoliation and then stored in an Ar glovebox. We tested the viability of this method by preparing two $\lesssim$5L flakes of \NPS{}, treating one with PMMA while the other was left out in air for one week. In Fig.~\ref{SampleDegXAS} we show XAS measurements taken on these two samples at the PEAXIS beamline pre and post a four hour beam exposure at the Ni L$_3$ edge in a specular scattering configuration. The sample coated in PMMA showed minimal changes to XAS spectra, while the sample without the PMMA protective coating showed significant changes and a nearly vanishing XAS intensity. 

In addition to a change in XAS spectra, we have also observed a degredation of RIXS spectra as a function of beam exposure time. After 11.5 hours of measurements, we began to observe a monotonic decrease in relative intensity between elastic line and \NiS{} multiplet excitations [Fig.~\ref{SampleDegRIXS}]. From T~=~0~m to T~=~210 m, an overall broadening of the spectra can be seen coupled with a near vanishing of higher energy loss (\eloss{} $\gtrsim$ 2.5 eV) charge transfer excitations.

\section{Fitting Procedure}
\subsection{Bulk Fitting and Peak Assignment}
We began our analysis by exploring the parameter space of the $d^8$ Tanabe-Sugano (TS) diagram within a single-ion model with octahedral (\Oh{}) symmetry, and subsequently, trigonal (\trig{}) symmetry \cite{tanabeAbsorptionSpectraComplex1954}. We thus minimized the difference between the calculated energies and those of peaks \textit{A} - \textit{F} for the bulk spectra. Fig.~\ref{Oh-D3d} shows how the excited state energy levels for divalent nickel change from their free-ion values as a function of \Oh{} and \trig{} crystal field (CF) splittings compared to \eloss{} values of fitted peaks. For this calculation, our model included the \Oh{} and \trig{} CF splitting terms, \Dq{} and $\delta$, and Slater-Condon parameters, \Fdd{}. \Dq{} was initially set to the \eloss{} value of the nominal \Dq{} line from the RIXS spectrum \cite{ghiringhelliNiOTestCase2005}, with a best fit value of 1.05~eV. Additionally, we found the best agreement with the fitted peaks with a 49\% reduction of \Fdd{} from its atomic Hartree-Fock value, and a range of \trig{} splitting between $\delta\sim$ -60 and -100 meV. This analysis showed that the peaks between \eloss{}$~\sim$~1~-~1.75 eV can be explained as excitations within the valence band of \Nid{} in the presence of a trigonally-distorted octrahedral complex.

\renewcommand{\thefigure}{SIII}
\begin{figure}[t]
  \setlength{\belowcaptionskip}{-18pt}
   \setlength{\abovecaptionskip}{1pt}
  \begin{center}
    \includegraphics[width = 1\linewidth]{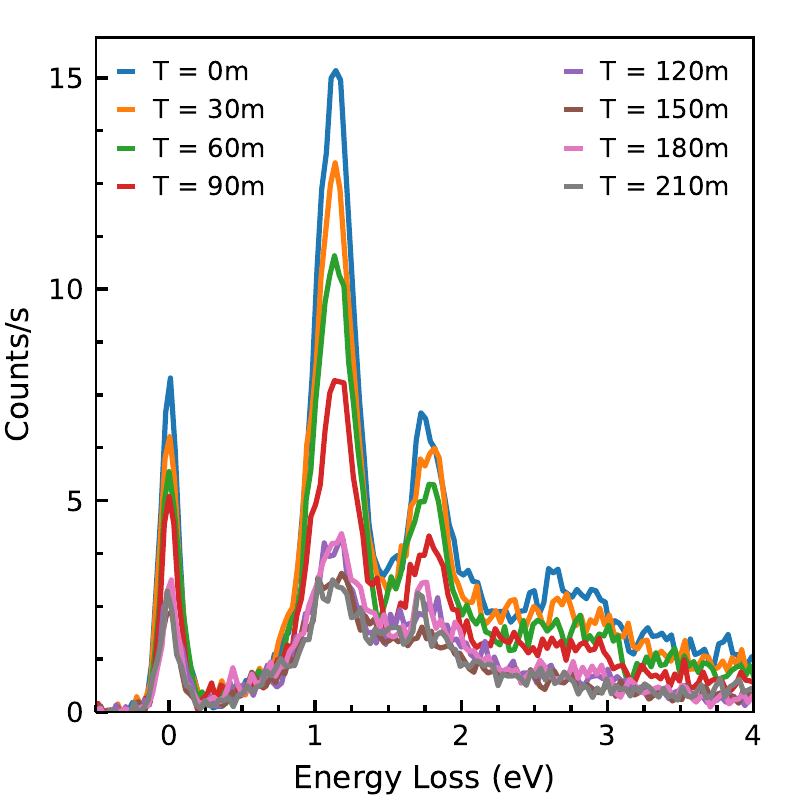}
  \caption{RIXS spectra at 160K at Ni L$_3$ resonance in 30 minute segments for an $\sim$ 5L samples of \NPS{}. RIXS spectra segments were collected after a previous 11.5 hours of beam exposure.} 
  \label{SampleDegRIXS}
  \end{center}
\end{figure}

\renewcommand{\thefigure}{SIV}
\begin{figure}[t]
  \setlength{\belowcaptionskip}{-18pt}
   \setlength{\abovecaptionskip}{0pt}
  \begin{center}
    \includegraphics[width = 1\linewidth]{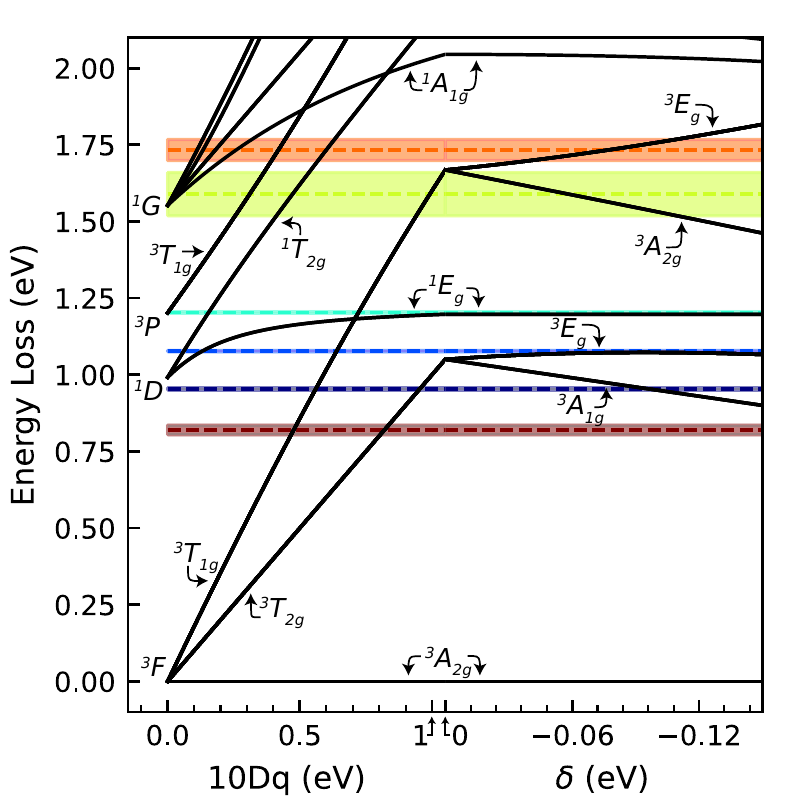}
  \caption{Calculated excited state energy levels for \Nid{} as a function of \Oh{} CF splitting \Dq{} (left) and \trig{} CF splitting $\delta$ (right) in the absence of spin-orbit coupling. \Fdd{} was fit to 49\% of its atomic Hartree-Fock value. Horizontal dashed lines and shaded regions show \eloss{} value of fitted peaks. Only symmetry labels relevant to the discussion in the main text are included.} 
  \label{Oh-D3d}
  \end{center}
\end{figure}

We now turn our attention to the determination of peak \textit{A}, which the single-ion model failed to capture. We repeated the above analysis as a function of \trig{} CF splitting within the multiplet ligand field (MLFM) model in order to understand the change from \Nid{} valence band excitations to full \NiS{} multiplet excitations, and thus, elucidate the origin of peak \textit{A}. Fig.~\ref{SIM-MLFM} shows the TS diagrams for these two models. Fig.~\ref{SIM-MLFM}(a) is equivalent to the rightmost panel in Fig.~\ref{Oh-D3d}, presented here including spin-orbit coupling. Fig.~\ref{SIM-MLFM}(b) shows the MLFM TS diagram using parameters discussed in the main text. Comparing peaks between \eloss{}$~\sim$~1~-~1.75 eV in these two TS diagrams shows how the introduction of additional interaction energies in the MFLM alter the \Nid{} valence band excitations. Including the analysis in Fig.~\ref{SIM-MLFM}(b), the complete minimization between calculated energies and peaks \textit{A}-\textit{F} is continued in Fig.~\ref{fig:BulkFitting}, where we explore the full range of \emph{free} MFLM parameters, including $\Delta$, \Dq{}, and the reduction factor to the Slater-Condon parameters \Fdd{}, denoted $F_{dd}$R. We find best agreement to our data with a small charge transfer energy, $\Delta$~=~0.60 eV, which gives rise to the low energy excitation present in Fig.\ref{SIM-MLFM}(b), and absent in Fig.\ref{SIM-MLFM}(a). We thus conclude that the origin of peak \textit{A} is a low energy charge transfer excitation and label it as \dCT{9}{1}. 

\subsection{3L Fitting}
As mentioned in the main text, empirically fitting the 3L RIXS spectra resulted in two scenarios of equally good fit quality with a minimum of six Gaussian peaks. Using physically meaningful parameters, these two scenarios can be summarized by the sign of the charge transfer energy ($\Delta$) extracted from modeling. Scenario one, where peak widths were fixed to the experimental resolution, resulted in a \emph{negative charge transfer energy}, while scenario two, where peak widths were allowed to relax, resulted in a \emph{positive charge transfer energy}. 

\subsection{Positive Charge Transfer Scenario}
Using physically meaningful parameters for octahedrally coordinated \NiS{}, we carried out a search of the parameter space for charge transfer energy $\Delta$, ligand-ligand and metal-ligand hoppings \Tpp{} and \pds{}, and trigonal distortion $\delta$, keeping $F(G)_{pd}$ fixed to 80\% of their Hartree-Fock values. The reduction factor for \Fdd{} was additionally kept fixed at the fitted bulk value. Fig.~\ref{fig:PCT_TS} shows the calculated energy of all excited states within a \NiS{} cluster as a function of these MLFM parameters shown with best fit values for $\Delta$ and trigonal distortion $\delta$, along with fixed hopping parameters for \Tpp{} and \pds{} obtained from DFT calculations.

\subsection{Negative Charge Transfer Scenario}
Following the same minimization procedure as the positive charge transfer scenario, Fig.~\ref{fig:NCT_TS} shows the calculated energy levels and best fit values for $\Delta$ and trigonal distortion $\delta$, along with fixed hopping parameters $T_{pp}$, $pd\sigma$. In addition to a negative charge transfer energy, $\Delta = -0.55$ eV, a reduced cubic crystal field splitting relative to bulk, 10\textit{Dq} = 0.31 eV, was also necessary to describe this fitting scenario.

\renewcommand{\thefigure}{SV}
\begin{figure}[t]
  \setlength{\belowcaptionskip}{-18pt}
   \setlength{\abovecaptionskip}{0pt}
  \begin{center}
    \includegraphics[width = 1\linewidth]{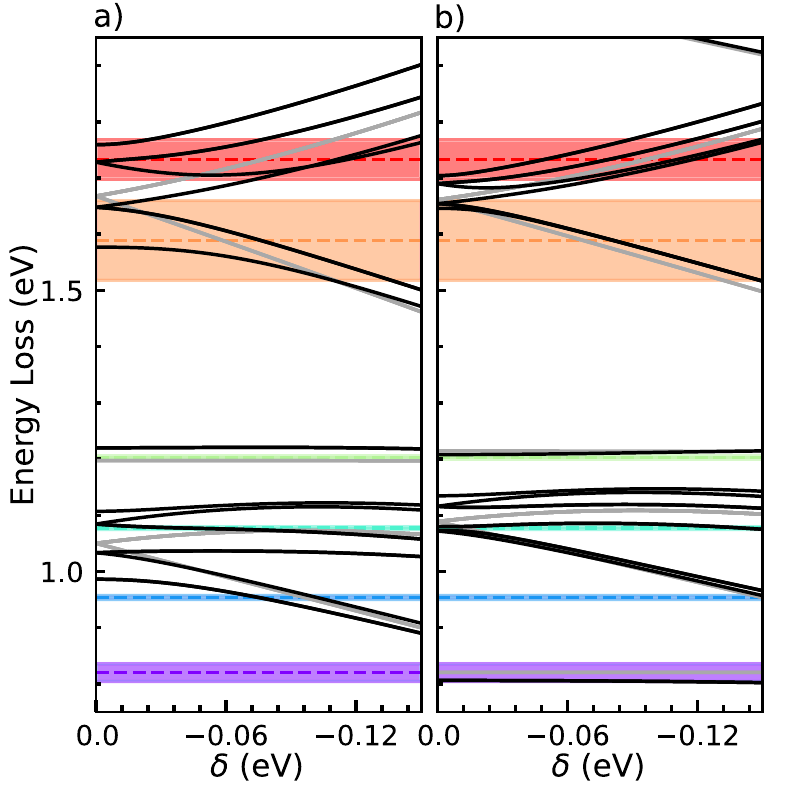}
  \caption{Calculated energy levels of excited states as a function of \trig{} CF splitting within a (a) single-ion model and (b) multiplet ligand-field model. Black and gray lines show energy levels calculated with and without SOC respectively. Horizontal dashed lines and shaded regions show \eloss{} value of fitted peaks } 
  \label{SIM-MLFM}
  \end{center}
\end{figure}

In order to facilitate a comprehensive search of the MFLM parameter space within fitting scenario one, we tracked the change in peak \textit{A} as a function of $\Delta$ vs. (\pds{}, \Tpp{}, $\delta$), shown as colormaps in Fig.~\ref{fig:Results_PeakAmaps}. Within the range of fixed hopping parameters for bulk and 1L \NPS{}, we find a clear delineation between the two fitting scenarios, where a peak \textit{A} position of $\sim$0.6 eV is only achievable with negative charge transfer values. 

\renewcommand{\thefigure}{SVI}
\begin{figure*}[]
  \begin{center}
  \setlength{\belowcaptionskip}{-18pt}
  \setlength{\abovecaptionskip}{0pt}
  \includegraphics[]{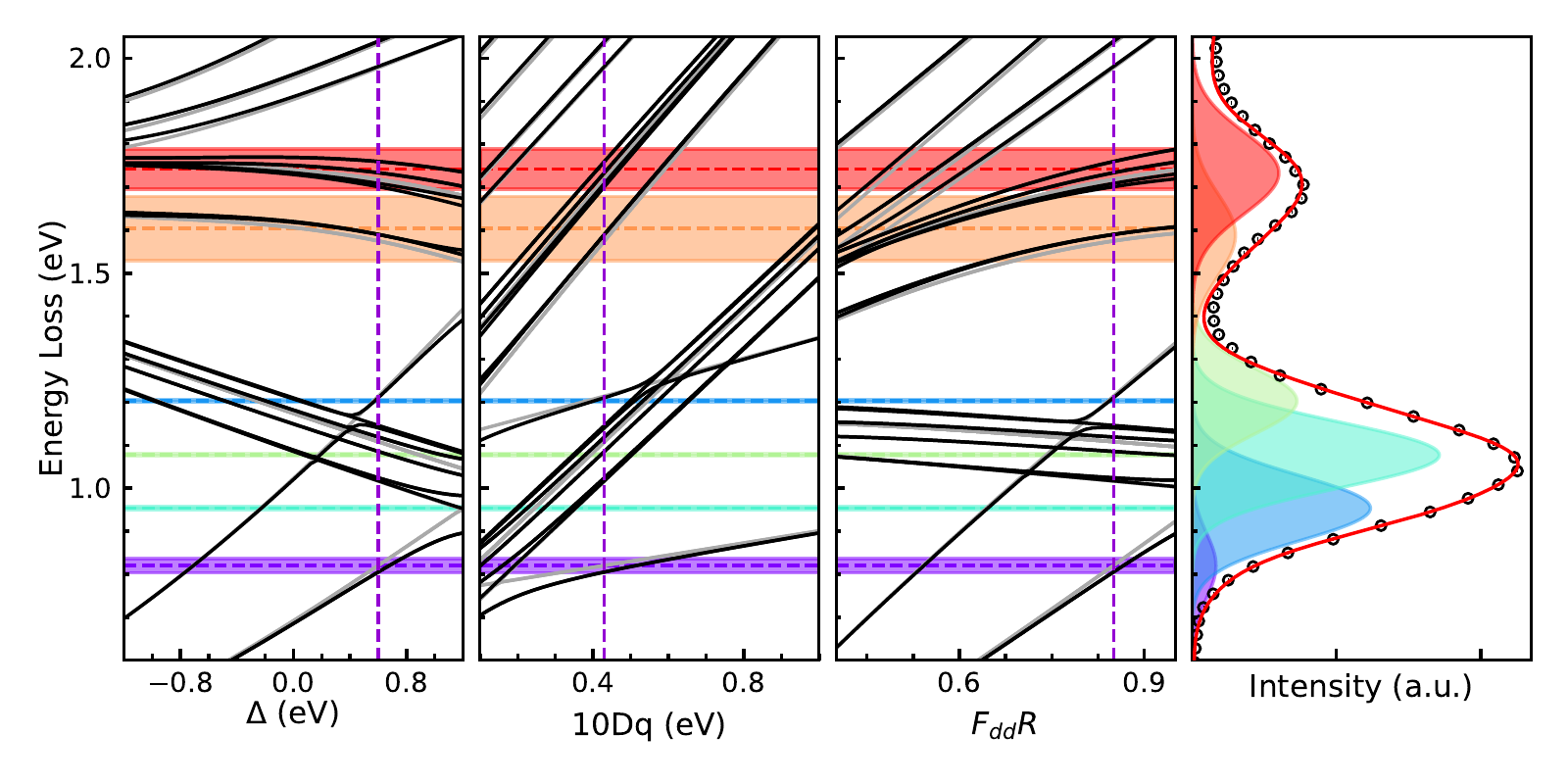}
  \caption{Calculated energy of all excited states as a function of MLFM parameters in comparison to bulk RIXS spectra. Horizontal dashed lines with respective shaded regions show the centroid position of fitted peaks. Vertical dashed lines represent the best fit values for each parameter.}
  \label{fig:BulkFitting}
  \end{center}
\end{figure*}

\renewcommand{\thefigure}{SVII}
\begin{figure*}[]
  \begin{center}
    \includegraphics[]{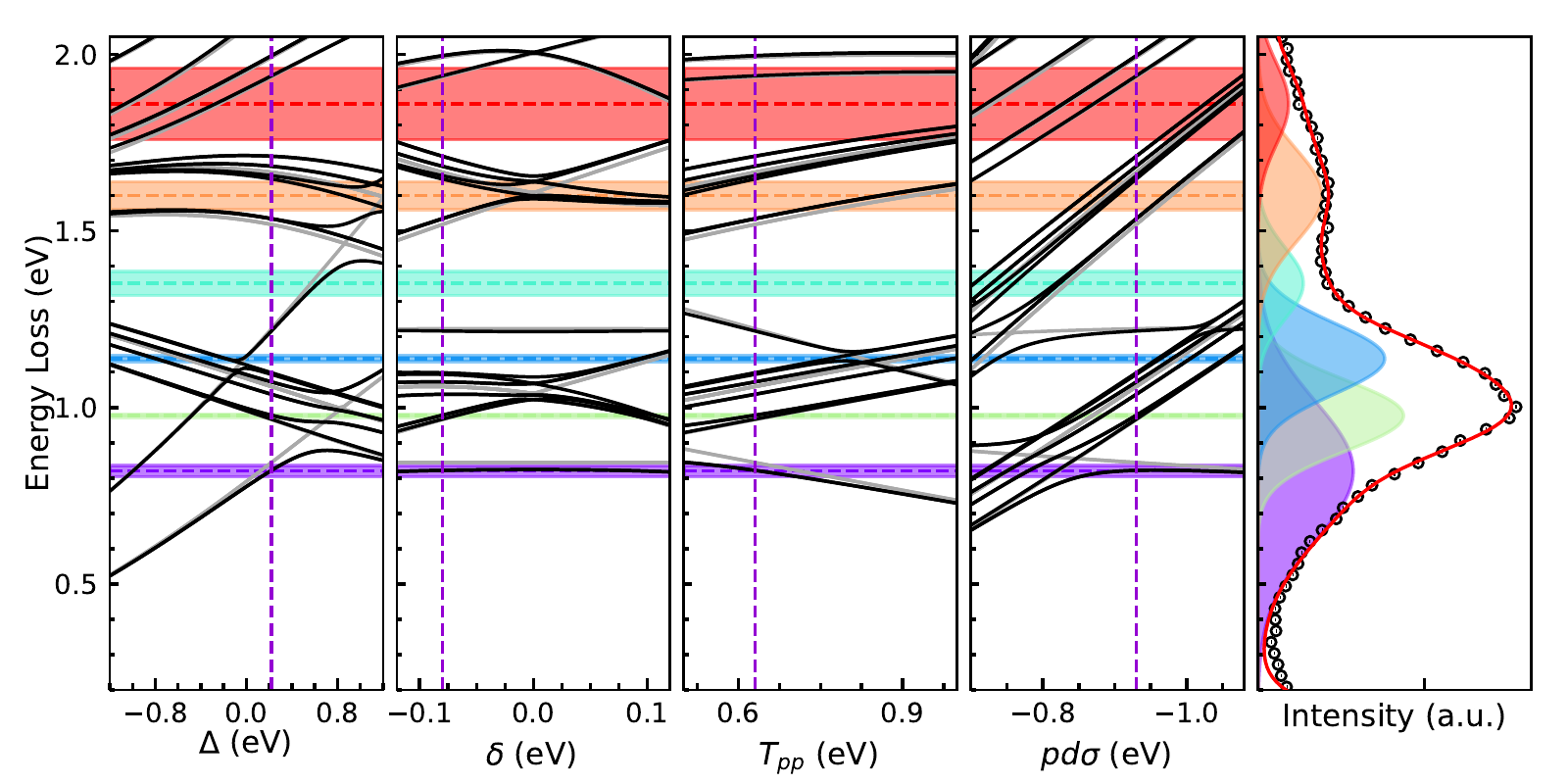}
  \caption{Calculated energy of all excited states as a function of MLFM parameters within fitting scenario two in comparison two 3L RIXS spectra. Horizontal dashed lines with respective shaded regions show the centroid position of fitted peaks. Vertical dashed lines represent the best fit values for each parameter.}
  \label{fig:PCT_TS}
  \end{center}
\end{figure*}

\renewcommand{\thefigure}{SVIII}
\begin{figure*}[]
  \begin{center}
  \setlength{\belowcaptionskip}{-15pt}
   \setlength{\abovecaptionskip}{0pt}
    \includegraphics[]{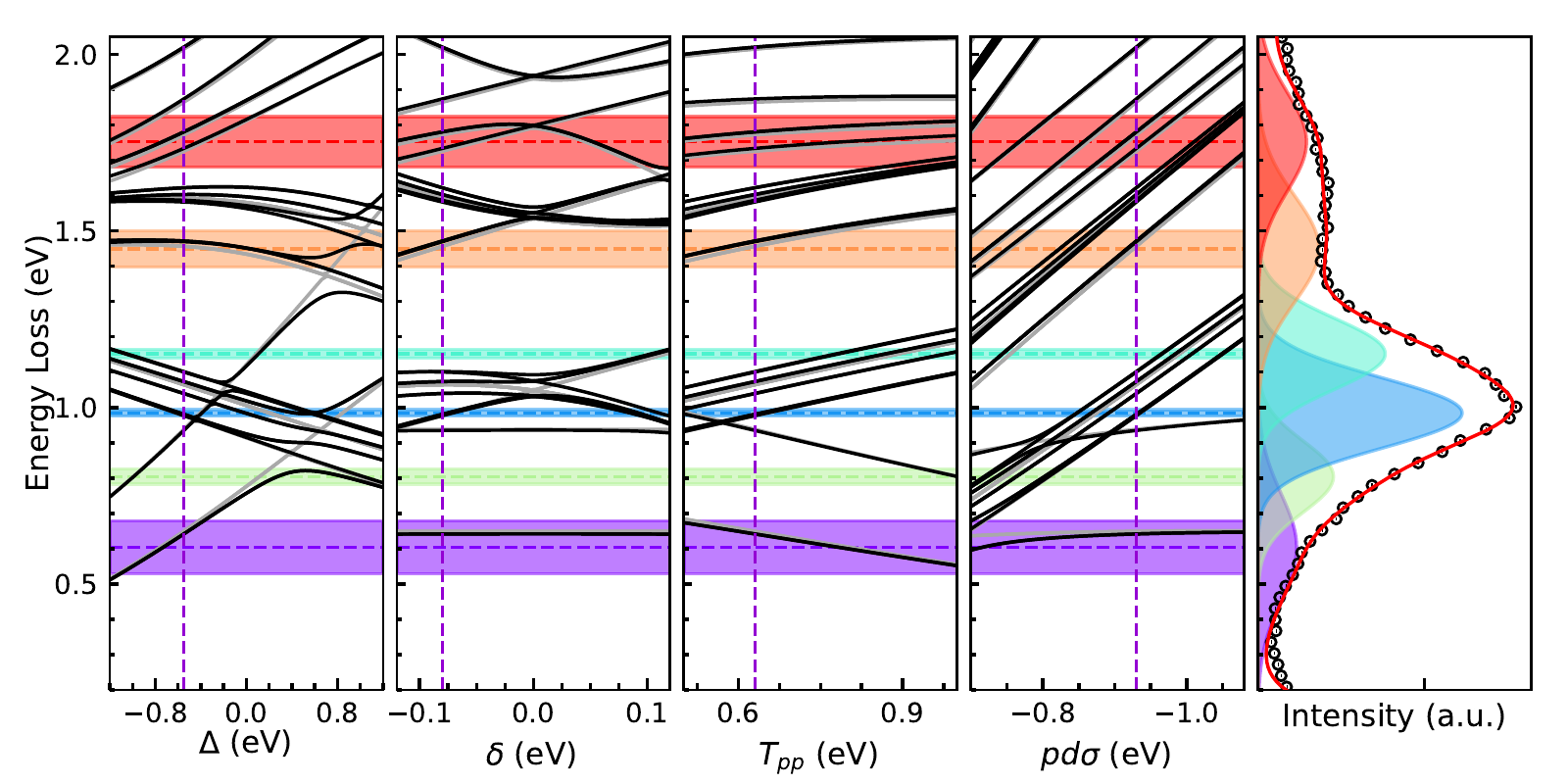}
  \caption{Calculated energy of all excited states as a function of MLFM parameters within fitting scenario one in comparison two 3L RIXS spectra. Horizontal dashed lines with respective shaded regions show the centroid position of fitted peaks. Vertical dashed lines represent the best fit values for each parameter.}
  \label{fig:NCT_TS}
  \end{center}
\end{figure*}

\renewcommand{\thefigure}{SIX}
\begin{figure*}[]
  \begin{center}
  \setlength{\belowcaptionskip}{-18pt}
   \setlength{\abovecaptionskip}{0pt}
    \includegraphics[]{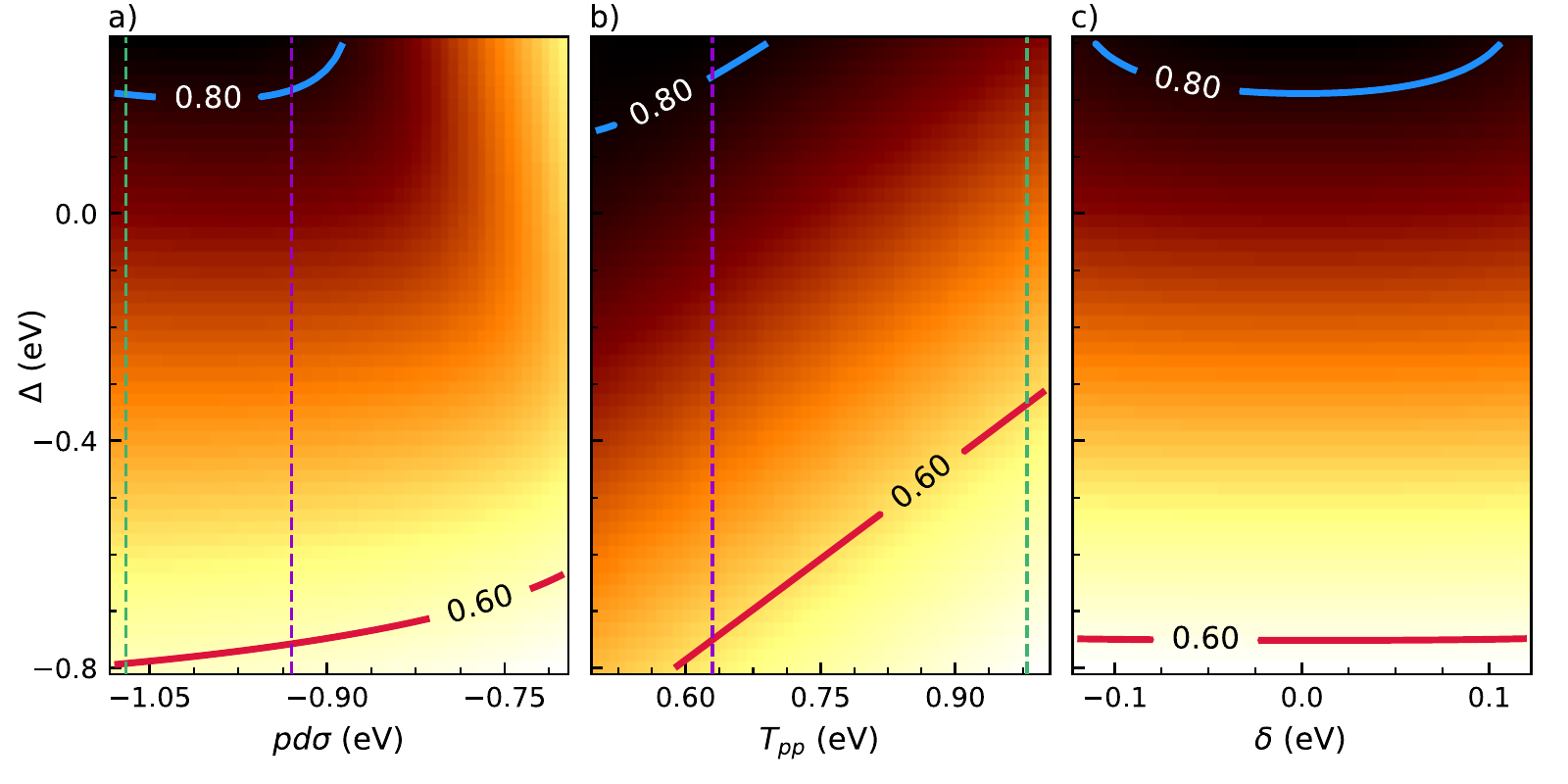}
  \caption{Calculated energy of \dCT{9}{1} excited state, peak \textit{A}, as a function of MLFM parameters. Contour lines indicate energies extracted from the two fitting routines for the 3L data, vertical dashed lines are parameter values from DFT for Bulk (violet) and 3L (green). These values are \pds{} = -0.93, -1.07, and \Tpp{} = 0.98, 0.63 for Bulk and 3L respectively.}
  \label{fig:Results_PeakAmaps}
  \end{center}
\end{figure*}

\section{Electronic Structure Details}

In this work, we used first principles simulations to accurately determine and constrain the ligand field theory parameters employed in the Anderson impurity model (AIM) used to model the experimental RIXS spectra. Our model which has a total of 14 adjustable parameters: $pd \sigma$, $pd \pi$, $pp \sigma$, $pp \pi$, charge transfer energy $\Delta$, spin-orbit coupling coefficients $\zeta_{d,i}$ of the initial state and $\zeta_{d,n}$ of the intermediate state, Coulomb energies $U_{dd}$ between $d$-orbitals and between $p/d$ orbitals $U_{dp}$, the reduction prefactors to the atomic Slater integrals $F_{dd}$, $F_{dp}$, $G_{dp}$ obtained from Hartree-Fock calculations of a free ion, crystal field splitting energy $10Dq$, and trigonal distortion $\delta$. 


\renewcommand{\thetable}{SII}
\begin{table}[h!]
\centering
\caption{Parameters and values used in MLFM calculations. Parameters not explicitly listed are the Ni 2\textit{p}-3\textit{d} on-site Coulomb interaction $U_{dp}$, fixed to 1.2$U_{dd}$, and the 3\textit{d} SOC for the initial and intermediate RIXS states, fixed to $\zeta_{3d,i} = 83$ and $\zeta_{3d,n} = 102$ meV respectively. $F_{dd}R$ and $F(G)_{pd}R$ are the Slater integral scalings applied to their atomic Hartree-Fock values. $F_{pd}$ and $G_{pd}$ are fixeed to 80\% of their atomic Hartree-Fock values.  }
\label{MFLMparams}
\begin{tabular}{||ccc||}
\hline
 \qquad Parameter \qquad & \qquad Bulk Model \qquad & \qquad 3L Model \qquad \\ \hline \hline
\textit{10Dq} & 0.43 eV & 0.40 eV \\ \hline
\textit{$\delta$} & -0.08 eV & -0.08 eV \\ \hline
\textit{U$_{dd}$} & 5.5 eV & 5.5 eV \\ \hline
\textit{$\Delta$} & 0.60 eV & 0.22 eV \\ \hline
\textit{\pds{}} & -1.07 eV & -0.93 eV \\ \hline
\textit{\pdp{}} & 0.67 eV & 0.46 eV \\ \hline
\textit{T$_{pp}$} & 0.98 eV & 0.63 eV \\ \hline
\textit{F$_{dd}$R} & 0.85 & 0.85 \\ \hline
\textit{\begin{tabular}[c]{@{}c@{}}F$_{dp}$R = \\ G$_{dp}$R\end{tabular}} & 0.8 & 0.8 \\ \hline
\end{tabular}
\end{table}


\subsection{Workflow for Determining Hopping Parameters from First Principles DFT Calculations} \label{supp-workflow}

We start with an outline of our approach for obtaining key impurity model parameters from first principles calculations since, to our knowledge, no one has explicitly documented a procedure for obtaining metal-ligand and ligand-ligand LCAO hopping parameters (e.g., $pd \sigma$, $pd \pi$, $T_{pp}$) from DFT in slightly distorted materials like NiPS$_3$. Originally, we aimed to employ a localization technique similar to Ref. ~\onlinecite{haverkortMultipletLigandfieldTheory2012} wherein we obtain atomic orbital-like Wannier functions which could be used in an LCAO/TB framework to calculate \NPS{}'s multiplet ligand field theory parameters. However, the slight trigonal distortion of \NPS{} and its more complex coordination environment compared to the ideal NiO octahedron of Ref.~\onlinecite{haverkortMultipletLigandfieldTheory2012} motivated our choice to instead use symmetry-adapted, maximally localized Wannier functions (MLWFs)\cite{sakumaSymmetryadaptedWannierFunctions2013, marzariMaximallyLocalizedGeneralized1997} since MLWFs are constructed directly from the converged DFT potential and do not require reconstruction of the potential between spherically symmetric local basis functions and interstitial plane wave basis functions as in the $N$MTO approach \cite{andersenMuffintinOrbitalsArbitrary2000}. Chronologically, our DFT/Wannier90 workflow consisted of: 

\begin{enumerate}
\item A self-consistent field calculation (described in further detail below) to solve for the set of single-particle orbitals (SPO's) and energies which converged the ground state charge density of nonmagnetic NiPS$_3$ at the PBE+$U$(=4 eV) level of theory, followed by;  
\item A non-self consistent mapping of the SPO's  onto a dense $6 \times 6 \times 6$ Monkhorst pack k-grid; 
\item Calculation of the band structure in the first Brillouin zone of the $P1$ bulk and monolayer structures to find the optimal band subspace $\left\lbrace \psi_{n \textbf{k}} \right\rbrace$ from which to obtain the MLWF's $w_n(\textbf{r}-\textbf{R})$,

\begin{equation}
    w_n(\textbf{r}-\textbf{R}) = \int_{BZ} \sum_m U^{\textbf{k}}_{mn} \psi_{m\textbf{k}}(\textbf{r})e^{-i\textbf{k}\cdot \textbf{r}}d\textbf{k}
\label{eqn:MLWF}
\end{equation}

\noindent
and corresponding unitary transformations $U_{mn}^{\textbf{k}}$. We chose the 27 isolated bands highlighted in Figure \ref{fig:band_subspace} as the subspace from which to obtain our 25 MLWF's since they contained almost all of the Ni $d$ and S $p$ partial density of states. Further details regarding the choice of NiPS$_3$ crystal structures and the DFT methodology are respectively provided in SI \ref{supp-struc} and SI \ref{supp-elec};

\renewcommand{\thefigure}{SX}
\begin{figure}[ht]
 \includegraphics[width=3.5 in]{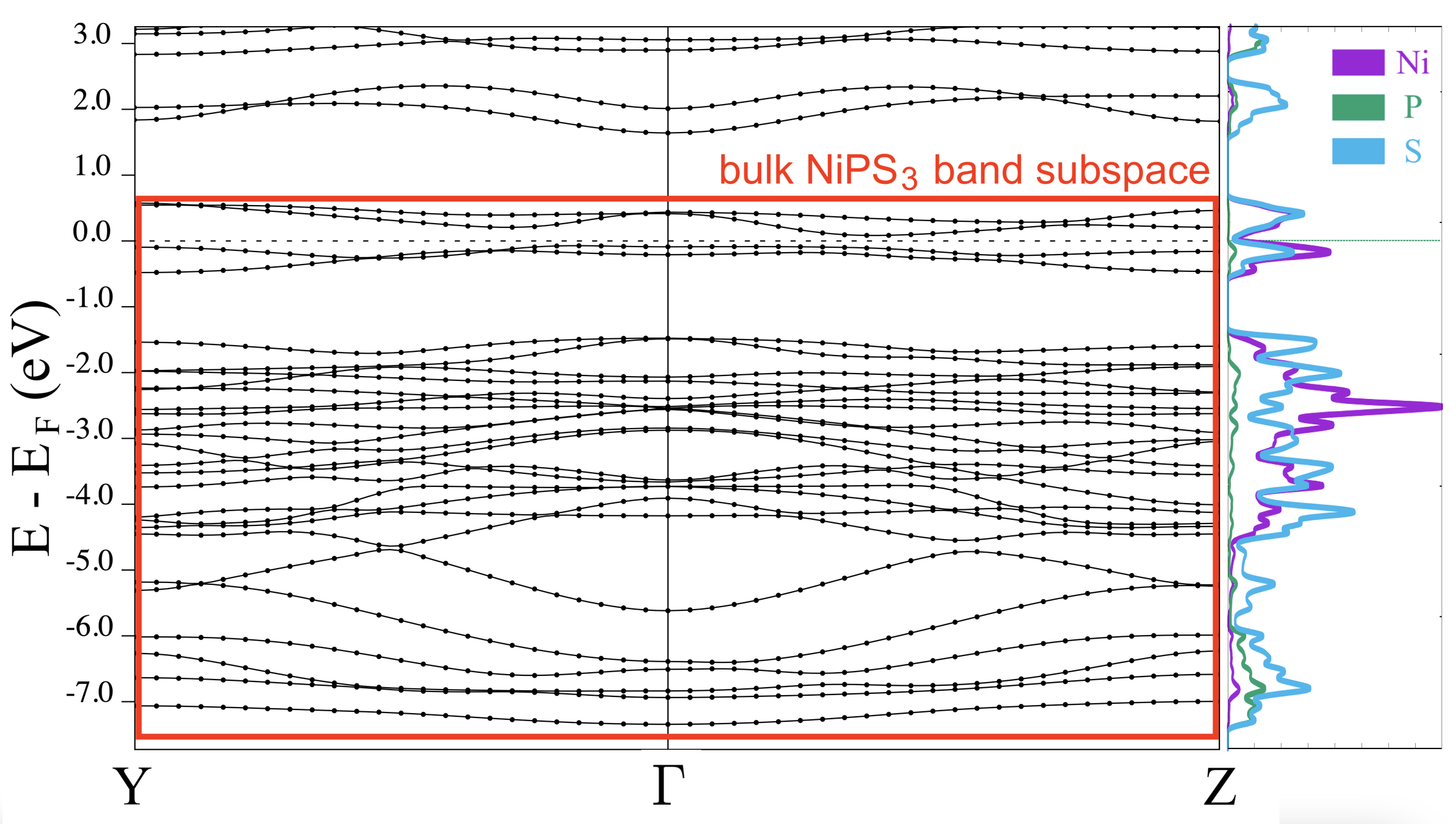}
 \caption{Bulk NiPS$_3$ band structure with the band subspace selected in this work outlined in red, and the corresponding partial density of states indicating the Ni $d$-orbital (purple), P $p$-orbital (green), and S $p$-orbital characters (blue) depicted on the right.
 }
 \label{fig:band_subspace}
\end{figure} 
\item Calculation of the initial overlaps of the Bloch-periodic parts of the SPO's; 
\item Conversion of the SPO's to Wannier90-readable formats using \textit{pw2wannier90} \cite{pizziWannier90CommunityCode2020}; 
\item Determination of the MLWF's according to Equation \ref{eqn:MLWF} \cite{marzariMaximallyLocalizedGeneralized1997};  and lastly
\item Calculation of metal-ligand and ligand-ligand hopping parameters as described hereafter.
\end{enumerate}

The Wannier90-obtained tight-binding Hamiltonian was a 25 $\times$ 25 matrix consisting of energies $E_{m,n} = \left\langle w_{m}(\textbf{r})|\hat{H}_1|w_{n}(\textbf{r}) \right\rangle$, where $m$ and $n$ respectively denote row and column indices and $w_{m}(\textbf{r})$ is the $m^{th}$ MLWF in the primitive cell. Our MLWF basis included the $d$-orbitals of both Ni atoms, the $p$-orbitals of each of the three distinct sulfurs coordinated to Ni1 (see Figure \ref{fig:structures}), and the $p$-orbitals of the two P atoms in our simulation cell, for completeness. The radial spreads of the Ni $d$-shaped MLWF's ranged from 0.6-1.1 \r{A} and those of the S and P $p$-shaped MLWF's ranged from 2-4 \AA, indicating physically reasonable localization. Further, the ratios of imaginary to real components of the MLWF's were 1 $\times$ 10$^{-6}$ or less, which indicated that the converged MLWF's were good-quality.

Next, we needed to select energy expressions from the overconstrained systems of LCAO equations presented in Ref. \onlinecite{slaterSimplifiedLCAOMethod1954} to solve for the metal-ligand hopping parameters. The presence of inequivalent sulfur sites and trigonal distortion in NiPS$_3$ yields many possible $pd\sigma$ and $pd\pi$ values upon solving for hopping parameters based upon different Ni-S pairs. To remove this ambiguity, we computed our $pd\sigma$ and $pd\pi$ values based upon the two largest energies $\left\langle w_{i}(\textbf{r})|\hat{H}|w_{j}(\textbf{r}) \right\rangle$ between Wannier function pairs based on the intuition that those with the largest energies contribute most significantly to the electronic structure of the material. Table \ref{tab:nips3_hop} provides the pairs of energies used to solve for the hopping parameters for each Ni-S pair in ML and bulk NiPS$_3$ where $d_b$ is the corresponding bond length (Å).



\renewcommand{\thetable}{SIII}
\begin{table}[ht]
\centering
\caption{Predicted NiPS$_3$ metal-ligand hopping parameters in eV and their corresponding largest energies.}
\hspace*{-0cm}\begin{tabular}{||c c c c c c||}
\hline
Sys & pair \; & $E_i$, $E_j$ & \; $pd \sigma$ & \; $pd \pi$ & $d_{b}$ (Å) \\ [0.5ex] 
 \hline\hline
\; bulk\;  & Ni1-S3 \; & $E_{x,xz}$, $E_{y,xy}$ & \; -1.07 & \; 0.81 & 2.384 \\ 
\hline 
\; bulk\;  & Ni1-S4 \; & $E_{z,z2}$, $E_{y,z2}$ & \; -0.92 & \; 0.77 & 2.379  \\ 
\hline
\; bulk\; & Ni1-S6 \; & $E_{x,x2y2}$, $E_{z,z2}$ & \; -1.23 & \; 0.43 & 2.400  \\ 
\hline
\hline
ML & Ni1-S3 \; & $E_{y,xy}$, $E_{x,z2}$  & \; -1.01 & \; 0.57 & 2.453  \\ 
\hline
ML & Ni1-S4 \; & $E_{x,xz}$, $E_{z,x2y2}$  & \; -1.08 & \; 0.32 & 2.452  \\ 
\hline
ML & Ni1-S6 \; & $E_{x,x2y2}$, $E_{z,z2}$  & \; -0.71 & \; 0.50 & 2.453 \\ 
\hline
\end{tabular}
\label{tab:nips3_hop}
\end{table}

As an example, solving for $pd\sigma$ and $pd\pi$ for the bulk Ni1-S3 pair entailed solving the system of equations 
\begin{gather}
E^{Ni-S}_{x,xz} = \sqrt{3} l_1^2 n_1 (pd \sigma)+n_1(1-2l_1^2)(pd \pi) \\
E^{Ni-S}_{y,xy} = \sqrt{3} m_1^2 l_1 (pd \sigma)+l_1(1-2 m_1^2)(pd \pi)
\end{gather}
\noindent
 where $l_1$, $m_1$, and $n_1$ are the direction cosines between the atomic Ni and S sites. We explicitly obtained symbolic expressions for the orbital combinations not considered in Ref. \cite{slaterSimplifiedLCAOMethod1954} using the compact closed-form treatment provided in Ref. \cite{podolskiyCompactExpressionAngular2004} and thereby verified their cyclic permutational symmetries\cite{slaterSimplifiedLCAOMethod1954}. The ligand-ligand hopping parameters $pp\sigma$ and $pp\pi$ (and thus $T_{pp} = pp\sigma - pp\pi$) were similarly determined via
\begin{gather}
E^{S-S}_{x,x} =  l_2^2 (pp \sigma)+(1-l_2^2)(pp \pi) \label{eq:pphop1} \\
E^{S-S}_{x,y} = l_2 m_2 (pp \sigma)- l_2 m_2 (pp \pi)
\label{eq:pphop2}
\end{gather}
\noindent
with the direction cosines $l_2$, $m_2$, and $n_2$ between the two S sites. Since only S3 and S4 were nearest neighbors, we report only one set of $pp \sigma$, $pp \pi$, and $T_{pp}$ parameters for the bulk and monolayer systems in the main text.

Additionally, it is noted that the same procedure using the bulk geometry (Table \ref{tab:both_relax}) with an added vertical vacuum of $>$ 25 Å to prevent interaction of images in the out-of-plane direction, `bulk+vac', yielded monolayer hopping parameters as reported in Table \ref{tab:nips3_hop2}. The resultant average of $pd\sigma$ and $pd\pi$ values
are respectively 0.29 eV and 0.06 eV larger in magnitude than those of the PBE-relaxed monolayer geometry `ML' (Table \ref{tab:both_relax}), which reflects the sensitivity of $pd\sigma$ and corresponding lack of sensitivity of $pd\pi$ to a Ni-S bond distance which is 0.06 Å shorter. Fits to the experimental data constrained by the unrelaxed monolayer `bulk+vac' hopping parameters were not robust whereas those contrained by the PBE-relaxed monolayer `ML' were, leading us to conclude that the change in RIXS response of NiPS$_3$ in the 2D limit is accompanied by only a slight increase in the 2D lattice constant relative to the bulk (0.5\%), and changes in $pd\sigma$ by 0.14 eV and $pd\pi$ by 0.21 eV.

\renewcommand{\thetable}{SIV}
\begin{table}[ht]
\centering
\caption{Predicted ``bulk" with vacuum NiPS$_3$ metal-ligand hopping parameters in eV and their corresponding largest energies.}
\hspace*{-0cm}\begin{tabular}{||c c c c c c||}
\hline
Sys & pair \; & $E_i$, $E_j$ & \; $pd \sigma$ & \; $pd \pi$ & $d_{b}$ (Å) \\ [0.5ex] 
 \hline\hline
\; bulk+vac\;  & Ni1-S3 \; & $E_{x,xz}$, $E_{y,xy}$ & \; -1.16 & \; 0.64 & 2.384 \\ 
\hline 
\; bulk+vac\;  & Ni1-S4 \; & $E_{z,z2}$, $E_{z,yz}$ & \; -0.96 & \; 0.62 & 2.379  \\ 
\hline
\; bulk+vac\; & Ni1-S6 \; & $E_{x,x2y2}$, $E_{z,z2}$ & \; -1.97 & \; 0.30 & 2.400  \\ 
\hline
\end{tabular}
\label{tab:nips3_hop2}
\end{table}

\subsection{Density Functional Theory Methodology} \label{supp-dft}

All simulations of structural and electronic properties were performed using DFT as implemented within the Quantum ESPRESSO package.\cite{giannozziQUANTUMESPRESSOModular2009} The PBE+$U$ functional with $U=4$ eV was selected to model this material since previous reports have shown this functional to yield structural parameters and band gaps that are consistent with experimental results.\cite{guNibasedTransitionMetal2019, kimMottMetalInsulatorTransitions2019} The key features of the real part of the experimental optical conductivity, $\sigma(\omega)$, of bulk (zigzag AFM) NiPS$_3$ are also reproduced when $\sigma(\omega)$ is calculated from DFT+$U$ using $U = 4$ eV.\cite{kimChargeSpinCorrelationVan2018} Further, a noncollinear Ne\'el antiferromagnetic (AFM) spin configuration was used for our relaxations, DFT parameter convergence, and preliminary band structure convergence calculations despite the experimentally observed zig-zag configuration of bulk NiPS$_3$, since experimentally comparable lattice parameters were recovered with Ne\'el ordering.\cite{wildesMagneticStructureQuasitwodimensional2015} Also, previous work, verified by our own calculations, demonstrates that the noncollinear Ne\'el antiferromagnetic and zigzag magnetic states are virtually degenerate for two-dimensional NiPS$_3$\cite{laneThicknessDependenceElectronic2020} and thus yield magnetic moments and band gaps with negligible differences. The final analysis for obtaining MLWF's and hopping parameters was however done for nonmagnetic NiPS$_3$ using the Neél-obtained plane-wave cutoff and k-grid parameters.

\subsubsection{Electronic Structure Analysis} \label{supp-elec}

To ensure that our DFT calculations were fully converged, first, we converged the Monkhorst-Pack k-point mesh and plane wave energy cutoff parameters for the PBE-relaxed bulk and monolayer structures (details may be found in SI \ref{supp-struc}) with the Neél-antiferromagnetic configuration, using the PBE+$U$ family of functionals.\cite{anisimovBandTheoryMott1991, ernzerhofAssessmentPerdewBurke1999} A k-point mesh of $6 \times 6 \times 6$ for the bulk and $4 \times 4 \times 2$ for the monolayer was sufficient to converge their respective total energies to within 5$\times 10^{-5}$ Ry. The plane-wave cutoff energy was converged to within chemical accuracy on the converged k-grid at a high cutoff of 600 Ry (bulk shown in Fig. \ref{fig:bulk_ecut}) despite the use of ultra-soft ccECP's designed for plane-wave calculations.\cite{annaberdiyevNewGenerationEffective2018} 

\renewcommand{\thefigure}{SXI}
\begin{figure}[ht]
 \includegraphics[width=2.8 in]{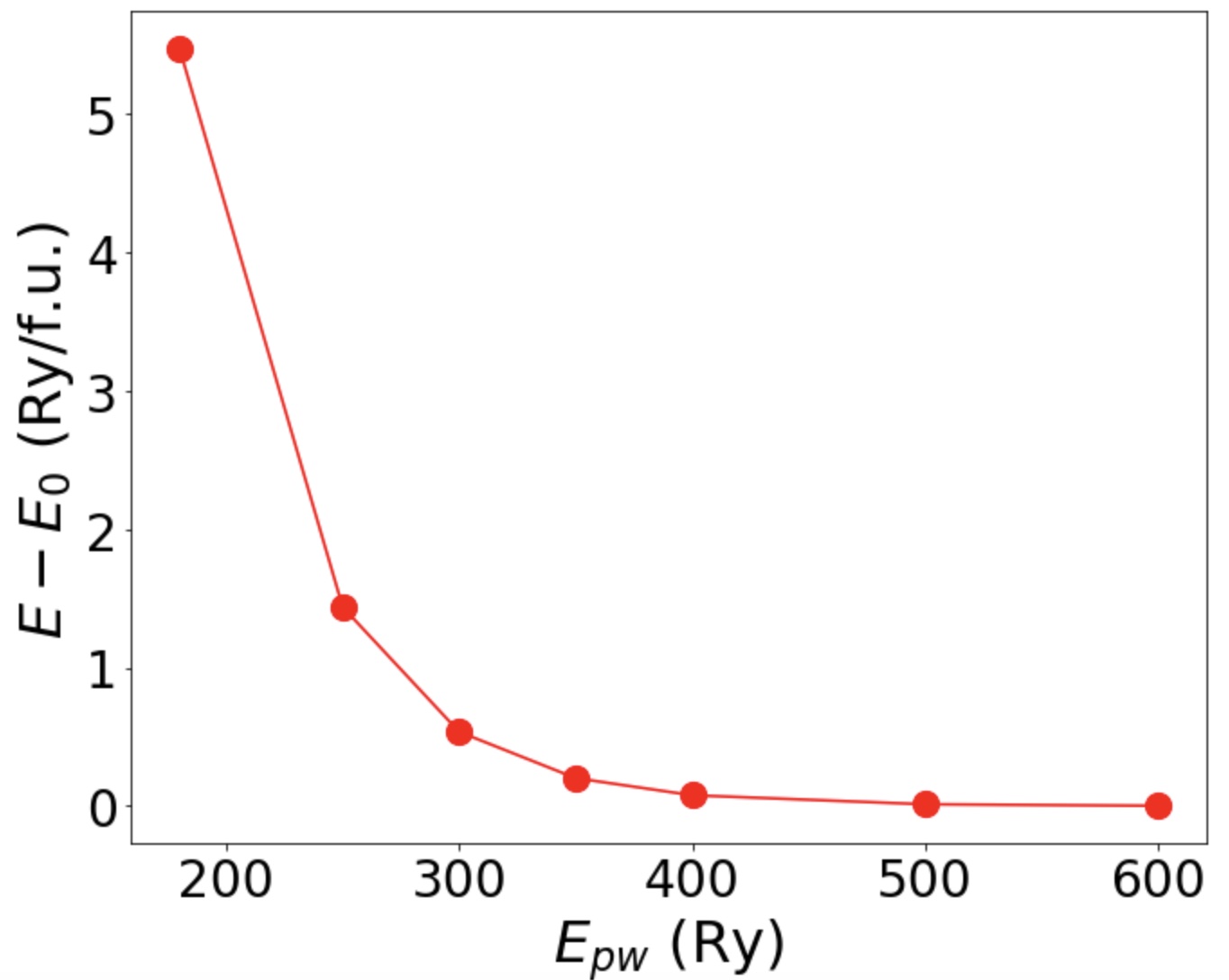}
 \caption{PBE+$U(=2$ eV$)$-calculated total energies ($8 \times 8 \times 8$) of bulk coordinate-relaxed NiPS$_3$ constrained to the experimental cell parameters of Ref. \onlinecite{wildesMagneticStructureQuasitwodimensional2015} as a function of the plane wave energy cutoff $E_{cut}^{PW}$.
 }
 \label{fig:bulk_ecut}
\end{figure}

Due to the steep computational expense expected to accompany non-self consistent field and band structure calculations of NiPS$_3$ with plane wave cutoffs of 600 Ry, we also performed tests to explicitly see how the band structure of bulk Ne\'el AFM NiPS$_3$ with the Materials Project ($C2/m$) structure is affected when it is calculated using partially converged plane wave cutoff energies. As illustrated in Figure \ref{fig:bulk_band_test}, a band structure obtained with a plane wave cutoff of 250 Ry is essentially identical to the $E_{cut} = 600$ Ry-obtained band structure. We thus deemed the subspace spanned by the wavefunctions obtained at $E_{cut} = 250$ Ry sufficient for obtaining accurate Wannier90 tight-binding parameters, and all of our Wannier90 analysis is done on wavefunctions obtained at this level of theory.

\subsubsection{Obtaining Bulk and Monolayer Crystal Structures} \label{supp-struc}

\renewcommand{\thefigure}{SXII}
\begin{figure}[btp]
\centering
 \includegraphics[width=3.41 in]{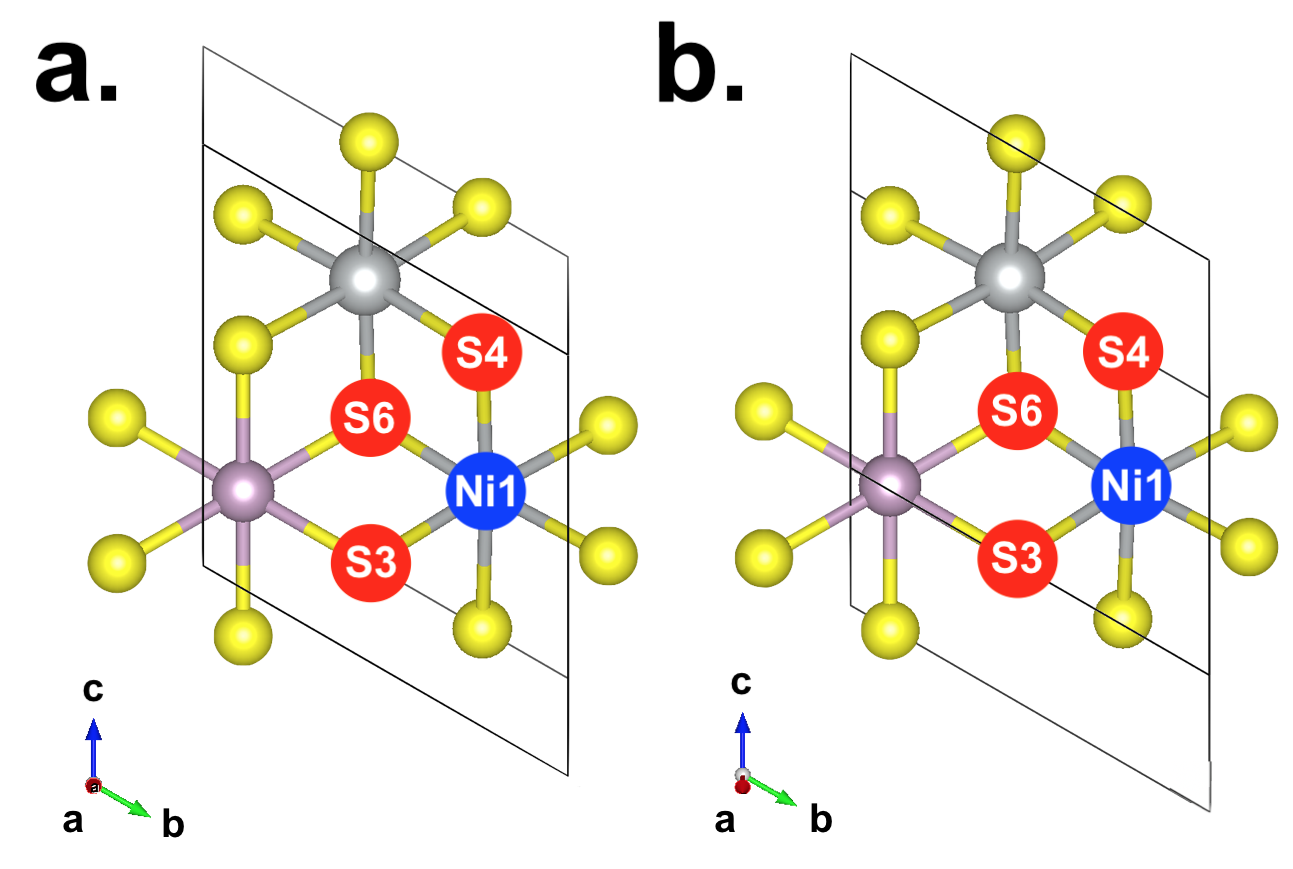}
 \caption{Top views of the crystal structures of (a) monolayer NiPS$_3$ optimized with PBE and (b) bulk NiPS$_3$ constrained to experimental lattice vectors\cite{wildesMagneticStructureQuasitwodimensional2015} with atomic coordinates using PBE+($U$=4 eV).} 
 \label{fig:structures}
\end{figure}

To obtain relaxed bulk and monolayer NiPS$_3$ crystal structures (see Figure \ref{fig:structures} and Table \ref{tab:both_relax}), the bulk structure with experimentally-determined lattice parameters having a monoclinic ($C2/m$) symmetry and a 10-atom unit cell consisting of two NiPS$_3$ formula units was first obtained from the Materials Project (MP) website \cite{jainCommentaryMaterialsProject2013}. The lattice vectors of the bulk structure were constrained to the experimental neutron diffraction values of crystalline NiPS$_3$ reported by Ref. \onlinecite{wildesMagneticStructureQuasitwodimensional2015} and atomic coordinates were relaxed using the Broyden-Fletcher-Goldfarb-Shanno quasi-Newton algorithm until ionic forces were smaller than $10^{-4}$ a.u. This yielded the bulk structure having $P1$ symmetry on which our DFT and Wannierization was performed (Figure \ref{fig:structures}a).
\renewcommand{\thetable}{SV}
\begin{table}[ht]
\centering
\caption{Relaxed NiPS$_3$ crystal structure parameters.}
\hspace*{-0cm}\begin{tabular}{||c c c c c c||}
\hline
 & $a$ (Å) & $b$ (Å) & $d_{Ni-Ni}$ (Å) & $\overline{d}_{Ni-S}$ & $d_{P-P}$ (Å) \\ [0.5ex] 
 \hline\hline
\; bulk \; \; & 5.8114 \; & 10.0640 & 3.3532 \; & 2.39(1) \; & 2.1848 \; \\ 
\hline 
\; ML \; \; & 5.8507 \; & 10.1347 & 3.3781 \; & 2.452(1) \; & 2.1927 \;  \\ 
\hline
\end{tabular}
\label{tab:both_relax}
\end{table}

One monolayer of NiPS$_3$ was cleaved from the starting MP structure and a vacuum of $>30$ \AA \, was added to prevent interlayer interactions between images for the monolayer simulations. Then, to inform our choice of exchange-correlation functional for monolayer variable-cell relaxation since no experimental monolayer lattice parameters exist, we performed variable-cell relaxations of bulk NiPS$_3$ in the collinear Neél AFM magnetic configuration starting from the MP structure using the LDA and PBE functionals with and without Hubbard $U = 2.0$ eV, which yielded the lattice parameters in Table \ref{tab:bulk_relax} and which are displayed alongside the experimental lattice parameters of paramagnetic bulk NiPS$_3$ obtained by Ref. \onlinecite{wildesMagneticStructureQuasitwodimensional2015}. All bulk structural relaxations were performed such that all lattice vectors and angles could vary under only the constraint of $C2/m$ cell symmetry; the PBE+$U$-converged k-point meshes were employed for all cell relaxations. Evidently, the cell relaxation using the PBE functional yielded the closest lattice parameters to experiment; this thus provided our rationale for using the PBE-optimized monolayer NiPS$_3$ crystal structure (having $P1$ symmetry) for our monolayer electronic structure and MLWF analysis (see Figure \ref{fig:structures}b). 

\renewcommand{\thetable}{SVI}
\begin{table}[h!]
\centering
\caption{Bulk NiPS$_3$ crystal structure parameters after full-cell relaxation with different DFT functionals, $V_{xc}$, compared with those from experiment. Where a $U$ is specified in the table, we employed $U=4$ eV.}
\hspace*{-0cm}\begin{tabular}{||c c c c c c||}
\hline
 & $V_{xc}$ & $a$ (Å) & $b$ (Å) & $c$ (Å) & $\beta$ (°) \; \; \\ [0.5ex] 
 \hline\hline
\; Exp.\cite{wildesMagneticStructureQuasitwodimensional2015} \; & -- \; & 5.811 \; & 10.064 \; & 6.896 \; & 106.22 \; \; \\ 
\hline 
\; DFT \; & PBE \; & 5.855 \; & 10.140 \; & 7.149 \; & 105.81 \; \; \\ 
\hline
\; DFT \; & PBE+$U$ \; & 5.906 \; & 10.231 \; & 7.223 \; & 105.79 \; \; \\ 
\hline
\; DFT \; & LDA \; & 5.727 \; & 9.929 \; & 6.262 \; & 107.64 \; \;  \\ 
\hline
\; DFT \; & LDA+$U$ \; & 5.734 \; & 9.934 \; & 6.408 \; & 107.25 \; \;  \\ 
\hline
\end{tabular}
\label{tab:bulk_relax}
\end{table}

\renewcommand{\thefigure}{SXIII}
\begin{figure*}[t!]
\centering
 \includegraphics[width=\linewidth]{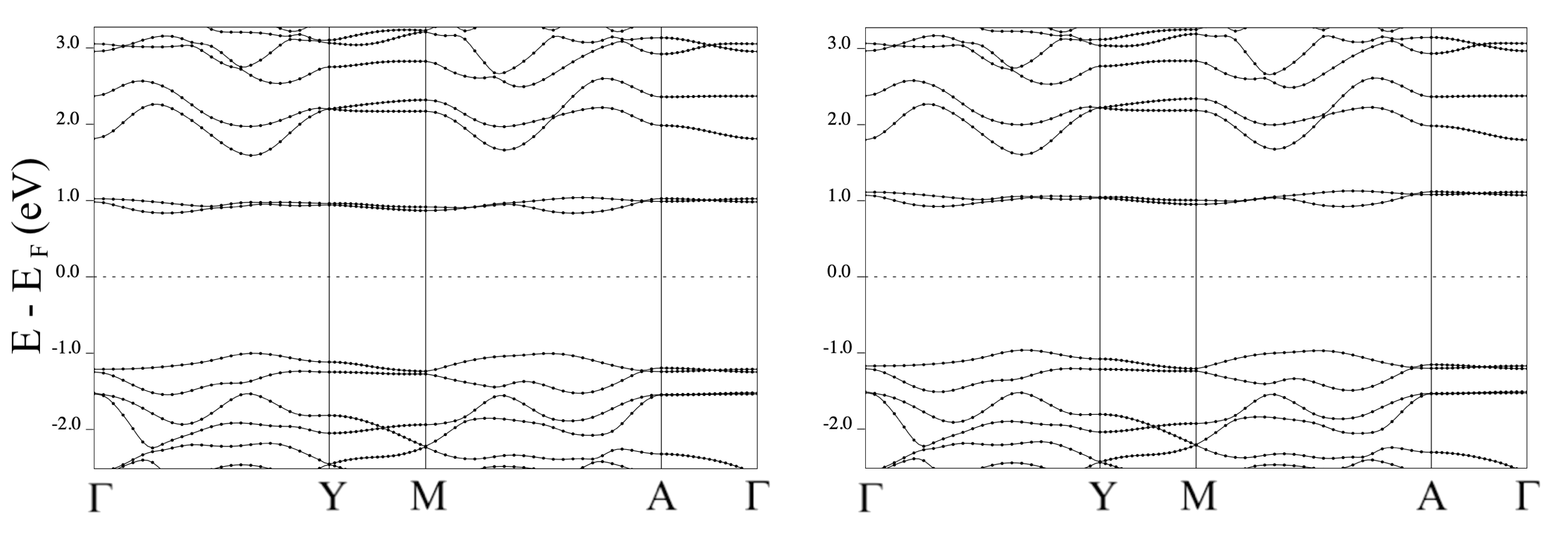}
 \caption{Band structures of Neél-AFM bulk NiPS$_3$ for PW cutoffs of (a) 250 Ry, and (b) 600 Ry (fully converged). Crystal structure was vc-relaxed with PBE+$U(=4 $ eV)}
 \label{fig:bulk_band_test}
\end{figure*}

\section{Superexchange Expressions}
The exact ground state wavefunction within an NiS$_6$ cluster is given by $\ket{\Psi_g} = \alpha\ket{3d^8} + \beta \ket{3d^9\underline{L}^1} + \gamma \ket{3d^{10}\underline{L}^2}$. Due to the magnitude of the trigonal field obtained from modeling ($\delta = -80$ meV), we simplify the calculation by approximating Ni-S bonds within an NiS$_6$ cluster as orthogonal. As mentioned in the main text, the ground state character for bulk and 3L \NPS{} finds a nearly equal mixture of $\ket{3d^8}$ and $\ket{3d^9\underline{L}^1}$ ($|\alpha|^2 \approx |\beta|^2$), with a negligibly small $\ket{3d^{10}\underline{L}^2}$ character. Thus, we independently apply the cell-perturbation to these two states. For fully occupied (inactive) $t_{2g}$ orbitals, $\ket{3d^8}$ and $\ket{3d^9 \underline{L}^1}$ are given by

\begin{equation}
\begin{split}
&\ket{3d^8}  =  \ket{d_{x^2-y^2}d_{3z^2-r^2}} \\
&\ket{3d^9\underline{L}^1} = \frac{1}{\sqrt{2}}(\ket{d_{x^2-y^2} L_{3z^2-r^2}} + \ket{L_{x^2-y^2}d_{x^2-y^2}}
\end{split}
\end{equation}
where $d_{x^2-y^2}$ and $d_{3z^2-r^2}$ denote holes in the two Ni 3\textit{d} $e_g$ orbitals, and $L_{x^2-y^2}$ and $L_{3z^2-r^2}$ denote holes in the S $3p$ molecular orbitals. These molecular orbitals are given by $L_{x^2-y^2} =$ \Lxy{} and $L_{3z^2-r^2} =$ \Lz{}, and \pir{i}{x,y,z} denote holes in the S 3\textit{p} orbitals. Utilizing the hopping parameters obtained from \textit{ab initio} calculations, we evaluate the two-center atomic overlap integrals within the Slater-Koster scheme of linear combinations of atomic orbitals. The superexchange for \de{} and \dn{} states are defined as \Jide{i} and \Jidn{i}, where \textit{i} $\in \{1,2,3\}$. 

The \iNN{1} superexchange is given by the fourth order perturbation term in \de{}, and second order term in \dn{}. These superexchange expressions are
\begin{equation}
\begin{split}
&J_1^{\alpha} = \frac{-\alpha^2 J_{H}^{S}(pd\sigma)^4}{\Delta''(\Delta''-J_{H}^{S}/2)(\Delta' + U_{pd})^2}
\\ &J_1^{\beta} = \frac{\beta^2}{2U_p} \biggl[\frac{1}{16}[(pp\sigma) + (pp\pi)]^2 + \frac{1}{144}[(pp\sigma)
\\ & \qquad \qquad \qquad \qquad \qquad \quad +9(pp\pi)]^2 - \frac{J_{H}^{S}}{9}\biggr] 
\end{split}
\end{equation}
where $\Delta' = \Delta + 2U_{pd}$, $\Delta'' = \Delta'  + U_{p}/2$, and $J_H^S$ is the Hund's coupling between S 3\textit{p} holes on a shared S site, assumed to be $\sim 0.7$ eV \cite{autieriLimitedFerromagneticInteractions2022}. $U_p$ is the on-site Coulomb interaction for S 3\textit{p} orbitals and is set to 4 eV \cite{babukaNewInsightStrong2017}. Note, here $U_{pd} = 1$ eV and is the Coulomb interaction term between S 3\textit{p} and Ni 3\textit{d} and is not the same as the $U_{pd}$ shown in the main text. For bulk \NPS{}, we find \Jide{1} = -1.8 meV and \Jidn{1} = -2.2 meV, then \Jide{1} = -1.6 meV and \Jidn{1} = -2.9 meV in 3L. 

The \iNN{2} superexchange is now given by the sixth order perturbation term in \de{}, and remains the second order perturbation term in \dn{}. These expressions are 
\begin{equation}
\begin{split}
&J_2^{\alpha} = \alpha^2 \Gamma \; (pp\pi)^2(pd\sigma)^4 \\
&J_2^{\beta} = \frac{\beta^2}{18U_p} (pp\pi)^2 \\
&\Gamma = \frac{2(\Delta' - U_{pd})^2 + \Delta'^2}{2(\Delta'-U_{pd})^4(\Delta'' - U_{pd}) \Delta'^2}
\end{split}
\end{equation}
These give \Jide{2} = 0.2 meV and \Jidn{2} = 0.052 meV for bulk, while \Jide{2} = 0.0021 meV and \Jidn{2} = 0.0003 meV for 3L \NPS{}.
Lastly, the \iNN{3} superexchange is again given by the sixth order perturbation term in \de{} and second order in \dn{}. We thus have 
\begin{equation}
\begin{split}
&J_3^{\alpha} = 2\alpha^2 \Gamma \; (pd\sigma)^4 \biggl[ \frac{1}{4} [(pp\sigma)-(pp\pi)]^2 + (pp\pi)^2 \biggr]
\\ &J_3^{\beta} = \frac{\beta^2}{24U_p}[(pp\sigma) - (pp\pi)]^2
\end{split}
\end{equation}
For bulk \NPS{}, these expressions give \Jide{3} = 12 meV and \Jidn{3} = 4.6 meV, while for 3L we find \Jide{3} = 8.6 meV and \Jidn{3} = 1.8 meV.




%

\end{document}